\documentstyle[aps,prb,manuscript,tighten,eqsecnum]{revtex} 
  
\begin{document} 
\title{Two-dimensional weakly interacting Fermi gas in a magnetic field:  Level
splitting} 
\author{Alexei Kolesnikov} 
\address{Ruhr-University Bochum, Universit\"atsstr. 150,\\ D-447801, 
Bochum, Germany\\ Landau Institute for
Theoretical Physics,\\ 
Kosygina St. 2, 117940, Moscow, Russia} 
\maketitle \begin{abstract} 
We apply the diagram-technique formalism beyond the Hartree-Fock 
approximation to a two-dimensional nearly ideal electron gas in a weak 
perpendicular magnetic field. The case of an almost completely filled 
upper Landau level (for its filling factor $\nu_0$ holds  $(1-\nu_0)\ll1$) 
with a quantum number 
$N_0 \gg 1$ is considered. We uncover two regimes of renormalization by 
electron-electron interactions. In the first regime, where $N^{1/2}_0(1-
\nu_0) \ll 1$, these interactions lead to a splitting of the Landau
levels. In the second regime, where $N^{1/2}_0(1-\nu_0) \gg 1$, apart of 
the splitting, a renormalization of the bare Zeeman splitting occurs. The 
intermediate case $N^{1/2}_0 (1-\nu_0)\approx 1$ cannot be studied within 
our approach. The applicability of the Fermi-liquid description is 
investigated for both regimes.  \end{abstract} 
 \section{Introduction} 

In the Fractional Quantum Hall Effect (FQHE), the Hall conductivity 
$\sigma_{xy}$, measured as a function of the magnetic field $H$, reveals 
plateaus at some fractional values of the filling factor \cite{FQHE}. A 
similar behavior of $\sigma_{xy}$ at integer filling factors is known as 
the Integer Quantum Hall Effect (IQHE) \cite{IQHE}. However, after the 
work of Laughlin \cite{Laug}, it became clear that in the FQHE, in contrast 
to the IQHE, electron-electron interactions play an important part. An 
essential advance in understanding of the FQHE was reached through the 
composite-fermion concept \cite{Jain}, further developed in Ref. \cite{HLR} 
for the filling factor $\nu=1/2$. There, the problem of a two-dimensional (2D) 
interacting electron gas in a strong perpendicular magnetic field was reduced 
to the investigation of composite fermions in an effective magnetic field. Since 
exactly at $\nu=1/2$ the field is gauged out, composite fermions were assumed to 
form a Fermi liquid. Such a system should, in an effective magnetic field, 
exhibit the de Haas-Shubnikov effect, the extreme form of which is the IQHE. 
Thus, the fractional Hall plateaus of the original electron system were 
interpreted as the integer ones of the transformed fermions.  However, the 
question crucial for such an approach, whether a 2D system of fermions in a 
magnetic field can indeed be treated in the framework of the usual Fermi-liquid 
theory, deserves a thorough investigation.   

The concept of the Fermi liquid makes it possible to pass from interacting 
fermions to non-interacting quasiparticles \cite{LAND}. The analysis in Refs. 
\cite{L,B-G} shows that interacting three-dimensional (3D) electrons in a 
sufficiently weak magnetic field can be described essentially as the ideal 
Fermi gas, after one changes all the properties of particles of the ideal Fermi 
gas to those of quasiparticles. On the other hand, the energy spectrum of a 
noninteracting 2D electron gas in a perpendicular magnetic field is known to 
consist of  discrete and dispersionless Landau levels (LL's). The 
renormalization of this spectrum should differ drastically from that in the 3D 
case and in systems with dispersion in the spectrum. The reason is that the LL's 
of noninteracting 2D electrons in a magnetic field are highly degenerate. 
According to the basic concepts of quantum mechanics, even weak interactions 
between the particles, can lift this degeneracy, sometimes resulting in 
nontrivial structures of the renormalized LL's.

Thus, the singularity in the density of states in 2D is reflected also in 
other physical quantities, whereas the singularity in 3D is smoothed because
of the presence of the third momentum along the direction of the magnetic 
field. More rigorously, according to Ref. \cite{L}, the self-energy part of 
the one-particle Green function $\Sigma$ depends on the magnetic field $H$ not 
only via the usual operator $\hat{{\bf  p}} -(e/c) {\bf  A} $, where ${\bf A}$ 
is the vector potential of the magnetic field, but also via a contribution of 
the order $(\hbar\omega_c/\mu)^{d/2}$, where $d$ is the spatial dimension of 
the system, the cyclotron frequency $\omega_c=eH/mc$, and $\mu$ is the Fermi 
energy. For $d=3$ and a weak magnetic field, the contribution of the second 
kind is small and could be disregarded in comparison with the first one, which 
is of the order of $\hbar\omega_c/\mu$. For a 2D system, both terms are
equally important and the applicability of the Fermi-liquid description
in a weak magnetic field is questionable. To check this, we will find all
essential contributions to $\Sigma$ and obtain the renormalized spectrum.

Some insight into the problems discussed above can be obtained with the
help of a rather simple model of interparticle interactions. In this paper,
the case of  short-range (contact) interactions is considered. An analogous 
analysis for such interactions was already performed for a 3D electron gas 
\cite{Gal}, for a 3D electron gas in a weak magnetic field  \cite{P-P}, and 
for a 2D electron gas \cite{Blood}. All these systems are examples of the 
traditional Landau Fermi liquid.

The question whether the Fermi-liquid theory is appropriate for the description 
of the 2D electron gas with contact interparticle interactions in a weak magnetic 
field, was investigated for the case of $N_0$ completely filled LL's \cite{WE}. 
The number $N_0$ is determined by the Fermi energy $\mu$, $N_0 = \mu/\hbar\omega_c 
\gg 1$. For a field $H$ (or density of electrons $n$), such that the upper LL $N_0$ 
is only partly filled, its filling factor $\nu_0 =nhc/2eHN_0 \lesssim 1$, the 
problem is still open. The case of incomplete filling is very interesting from
the experimental point of view as well as for theoretical study. In the present 
work, we investigate how a deviation of $\nu_0$ from unity changes the spectrum of 
the interacting system, treating $1-\nu_0$ as a small parameter. 

We find the spectrum for two regimes. In the first regime, where $N^{1/2}_0(1-\nu_0) 
\ll 1$, a splitting of the LL's occurs. In the second regime, where $1/N^{1/2}_0 \ll 
1-\nu_0 \ll 1$, the splitting is accompanied by a shift of the Zeeman-split LL's. 
The intermediate case $N^{1/2}_0 (1-\nu_0)\approx 1$ is the most complex and we 
were not able to study it. 

The renormalization of the spectrum has an unusual character, compared with all 
known cases. Such features of the bare spectrum, as the absence of dispersion and 
the equal spacing between the neighbouring Landau levels, lead to resonances which 
{\it split} the bare spectrum.  The renormalization of the Zeeman-split LL's by 
long-range Coulomb interactions is a well-known effect for strong \cite{Ando} and 
intermediate \cite{MAC} magnetic fields. This result is of a similar origin with the 
second regime, being a manifestation of the asymmetry in interactions of particles 
having opposite spin projections with the electrons from the 
Fermi sphere. 

It is noteworthy that there is a certain similarity between the renormalization in 
our system and that of fermion systems with a dispersion in the bare spectrum: We 
will find criteria which determine when the system under consideration belongs to 
Fermi liquids. They provide conditions for the bare spectrum near the LL $N_0$ to 
be weakly renormalized, so that classification of the LL's according to their 
indices is preserved, i.e. the LL's do not overlap with each other. 

The spectrum obtained shows that electron-electron interactions do not break the 
translational symmetry and thus partially preserve the initial degeneracy of the 
spectrum. In contrast, introducing impurities into the sample leads to a broadening 
of the bare LL's and the formation of bands of delocalized states \cite{???}.

It is worth mentioning that in our work, we study the spectrum of 
quasiparticle-type excitations whose physical properties are strongly affected by  
electron-electron interactions. Quantities like the heat capacity and the spin
paramagnetism are sensitive to it. In contrast, the cyclotron effective mass and 
the zero-field plasma frequency are not changed by these interactions \cite{Kohn}.

This paper is arranged as follows. In Section II, we introduce the model. In
Section III, we detect the effect of splitting, crucial for our system, within 
second order of perturbation  theory. Then, with the aid of general rules 
formulated in Section IV, we select important diagrams for the LL's in Section 
V. In Sections VI and VII, respectively, the renormalization of LL's near and 
further away from the LL  $N_0$ is studied. Section VIII contains the summary. 

\section{Nearly ideal Fermi gas}

In this work we study the properties of a 2D system of electrons in a uniform
magnetic field. We consider the Hamiltonian \begin{eqnarray}
\hat{H}=\hat{H}_{0}+\hat{H}_{\rm int} \; ,\end{eqnarray} where the
Hamiltonian of the noninteracting electrons (apart from the Zeeman term), 
$\hat{H}_{0}$, is given by \begin{eqnarray} \hat{H}_{0}=\frac{1}{2m} \,\int 
{{\rm d} {\bf r}}\,\hat{\psi}^+_\sigma({\bf r})\, \left[ {-i \hbar{\bf 
\nabla}+e{\bf A}({\bf r})/c} \right]^2\,\hat{\psi}_\sigma({\bf r}) \label{a2} 
\; . \end{eqnarray} In Eq. (\ref{a2}), summation over the spin quantum 
number $\sigma$ is implied, $\hat{\psi}^+_\sigma({\bf r})$ and 
$\hat{\psi}_\sigma({\bf r})$ are the creation and annihilation operators, 
${\bf A}({\bf r})$ is the vector potential of an external magnetic field $H$; 
$m$ is the effective mass of the electron and $-e$ is its charge.

We use the term ``nearly ideal" for a Fermi gas in which the short-range  
repulsive pair-interaction potential $U({\bf r})$ with range $a$ simultaneously 
satisfies two conditions \begin{eqnarray} ak_f \ll 1 \label{a3}  \; , \quad 
\frac{mU_0}{\hbar^2}\, \ln\left(\frac{1}{ak_f}\right) \ll  1 \; . \end{eqnarray} 
To disregard corrections to the bare vertex from the Cooper chanel in the case of 
a weak magnetic field, more severe condition $mU_0/\hbar^2\ln\left(N_0
\right)\ll 1$ should be imposed on the potential $U({\bf r})$. In Eq. (\ref{a3}),
 $U_0=\int U({\bf r})\, {\rm d} {\bf r}$ and $p_f=\hbar k_f$ is the Fermi 
momentum. The first of conditions (\ref{a3}) means that one has a dilute system 
and the second assures the applicability of the Born approximation in 2D. Because 
of  the short-range nature of the potential, only interactions between particles 
with opposite spins are important: \begin{eqnarray} \hat{H}_{\rm int}=U_0  \int 
{\rm d}  {\bf r} \, \hat{\psi}^+_\uparrow  ({\bf r}) \, \hat{\psi}^+_\downarrow 
({\bf r}) \,\hat{\psi}_\downarrow ({\bf r}) \, \hat{\psi}_\uparrow({\bf r}) 
\label{a4} \; . \end{eqnarray}

We adopt the Landau gauge for which the single-particle wave functions
are characterized by two quantum numbers, the number $j$ of the LL and one
(conserved) projection of the momentum $p$: \begin{eqnarray}
\phi_{jp}({r_x,r_y})= (2^jj! \sqrt{\pi})^{-1/2}\,e^{ipr_y}\,
e^{-(r_x+p)^2/2}\, H_j(r_x+p) \label{a5} \; , \end{eqnarray}
where $H_n$ is a Hermite polynomial, $l_H=\sqrt{\hbar c/eH}$ is the 
magnetic length. Hereinafter, the dimensionless variables $r_x/l_H$, $pl_H/\hbar$ 
will be used.

We employ the temperature  diagram-technique  formalism, which is due to Gibbs 
averaging capable of dealing with systems with degeneracy \cite{AGD}. The 
single-particle Green function in a coordinate representation, expanded with respect 
to the basis $\phi_{jp}$, is \begin{eqnarray}G_0({\bf r},{\bf r}^\prime;\omega_k)=
\sum_{j,\, p,\,\sigma}\frac{\phi_{jp}({\bf r})\phi^*_{jp}({\bf r}^\prime)}
{i\omega_k-\epsilon_{j, \,\sigma}} \label{a6} \; , \end{eqnarray} where $\omega_k=
\pi T(2k+1)$ is a fermion Matsubara frequency. The bare one-particle energies 
are the energies of the LL's $\epsilon_{j,\,\sigma}= \hbar\omega_c(j+1/2)+g 
\mu_B \sigma$ with the Bohr magneton $\mu_B= e\hbar /2m_0c$ and $m_0$ is the free 
mass of the electron. As it will become clear in the following, the features of the
spectrum we will obtain do not depend on the exact value of 
the $g$ factor as long as $g \neq 0, 2$.

In the following, we will sometimes omit the spin index $\sigma= \pm 1/2$  keeping 
only the number of the LL. For example, for the case of complete filling 
($\nu_0= 1$), each spin-split LL is either filled or not, so there 
exists a spin symmetry. It is obvious that in this case, all the effects of 
electron-electron interactions cannot depend on spin either. The dependence on 
spin will be given explicitly for all equations where it is important. In this 
case, the notation $N^\uparrow_0+j$ and $N^\downarrow_0+j$ will be used for LL's 
with the level index $N_0+j$  and the respective quantum number of spin 
$\sigma=1/2$ and $\sigma=-1/2$. We denote by $N^\uparrow_0$ the incompletely
filled LL corresponding to the chemical potential, suggesting that the LL 
$N^\downarrow_0$ and all LL's below it become completely filled in the
limit $T \rightarrow 0$.

After we have introduced $\hat{H}_{\rm int}$ and $G_0$, we proceed with the diagram 
technique in the standard manner \cite{AGD}. Expanding the annihilation 
and creation operators in Eq. (\ref{a4}) in functions $\phi_{jp}({\bf r})$, we can 
analyze the interaction of the electrons in terms of their transitions between 
of the LL's (cf. Fig. 1). It is convenient to redefine $G_0$ and $\hat{H}_{\rm 
int}$ attributing all four functions $\phi_{jp}({\bf r})$ appearing in
$\hat{H}_{\rm int}$ to the interaction vertex \cite{L-L}. As a result, only 
the denominators are left in sum (\ref{a6}). 
Conditions (\ref{a3}) assure the applicability of the Born approximation for a dilute 
2D system of electrons without magnetic field \cite{Blood}, with the vertex of 
order $V_0 \sim (mU_0/\hbar^2) p^2_f/m$. The presence of a magnetic field 
immediately changes the vertex. After switching on a weak magnetic field, an 
additional characteristic length $l_H$ is introduced such that $a \ll 1/k_f 
\ll l_H$. The non-interacting electrons in a magnetic field are described by 
the functions (\ref{a5}), four of which enter in the vertex of interaction. We 
consider the contact interaction, $U({\bf r}) \sim \delta({\bf r})$, so that 
$\phi^4({\bf r})$ is replaced with $\phi^4(0)$, and we arrive at the following
estimate for the vertex in a magnetic field: \begin{eqnarray} V \approx 
(m U_0/\hbar^2)\, (\hbar \omega_c) \label{a7} \; .\end{eqnarray} The parameter of 
perturbation theory  is $V/\hbar \omega_c \ll 1$. 

The spectrum of  one-particle excitations is determined by the poles of the 
of the analytical continuation of the temperature Green function $G^{-1}(\omega)
=G^{-1}_0(\omega)-\Sigma(\omega)$. We will obtain the self-energy part of the 
one-particle Green function $\Sigma(\omega)$  by summing up the series of the 
most important diagrams $\Sigma^{(n)}$. In the following sections, we derive the 
series by selecting the most essential contributions among all the diagrams 
arising in a given order $n$. In order to find out whether the Fermi-liquid 
description is correct for the system under consideration, we derive the expression 
for $\Sigma(\omega)$ as an expansion in small parameters, $1-\nu_0$, $1/N_0$ 
and $V/\hbar \omega$. We restrict ourselves to determining the 
dependencies of the diagrams on powers of these parameters, disregarding  
numerical coefficients and weaker functions of the parameters. In this way, 
we derive the Dyson equation and from it the spectrum. It is worth emphasizing,
however, that in spite of the approximations made in estimating of $\Sigma^{(n)}$,
this procedure is non-perturbative since the initial spectrum is degenerate
(cf. Conclusions).

\section{Second order of perturbation theory}

Any system of fermions belongs to the class of Fermi-liquid systems if near 
the Fermi momentum $p_f$ its Green function is given by \begin{eqnarray}
G(\omega)=Z/(\omega-\frac{p_f}{m^*}|p_f-p|) \label{b1} \; , \end{eqnarray}
where $m^*$ and $Z$ are the renormalized effective mass and the discontinuity 
of the momentum distribution of the particles at the Fermi surface,  
respectively. In Eq. (\ref{b1}), $\omega$ is measured from the chemical potential $\mu$. 
For a 2D or 3D Fermi gas without magnetic field or a 3D Fermi gas with 
magnetic field and  contact interaction, the first-order diagram gives rise 
to a trivial shift of $\mu$. The parameters of the Fermi-liquid 
theory, $Z$ and $m^*/m$, are determined by the second-order diagram (Fig. 2). 
They differ from those for the ideal Fermi gas $Z=1$, $m^*/m=1$ in second order 
of the parameter $(mU_0/\hbar^2)$, according to Ref. \cite{La-Li}.

Similarly, in the case of a 2D Fermi gas in a magnetic field, the first-order 
diagram gives only a constant contribution to the self-energy $\Sigma$ and 
simply shifts the chemical potential $\mu$. We shall disregard contributions 
of this kind, taking for $\mu$ its shifted value. In contrast, the 
second-order diagram  leads to a renormalization which drastically 
differs from the one in the absence of a magnetic field. In order to 
understand this effect and obtain the characteristic values of the parameters 
involved, let us start from the case of $H=0$ and then recover the 
corresponding result in a finite magnetic field. 

The second-order diagram determines the probability of a particle with momentum $p$
to decay. Its value at $H=0$ (all notations are introduced in Fig. 2) is given by 
\cite{AGD}  \begin{eqnarray} \Sigma^{(2)}(\omega, {\bf p}) \propto U^2_0  
\int{\frac{{\rm d} {\bf p_1}\, {\rm d}{\bf p_2}\, {\rm d}{\bf p_3}} {(2\pi
\hbar)^4}}\,\delta(\omega +\epsilon_1-\epsilon_2-\bf \epsilon_3)\,\delta({\bf p}+
{\bf p_1}-{\bf 
p_2} -{\bf p_3}) \label{b2} \; , \end{eqnarray} where the $\delta$-functions account 
for energy and momentum conservations. One writes $\delta(r)= \int \exp(iqr)\, {\rm 
d}q /2\pi$, integrates over the angles in $r$- and $p_i$-spaces, and obtains four 
Bessel functions using the representation \cite{Manual} $\int^{2\pi}_{0} \exp(i 
\alpha\sin \varphi)\,{\rm d} \varphi/2\pi=J_0(\alpha)$. The momenta $p$, $p_i$ in 
the magnetic field are given by ${\bf p^2_i}=2m\hbar\omega_c(N_i+1/2)$, so that the 
integration over ${\bf p^2_i}$ becomes a summation $\int{\rm d}{\bf p^2_i} 
\rightarrow \sum_{N_i} 2m\hbar\omega_c$ in the limit of a weak magnetic field, 
$N_0 \gg 1$. One obtains the second-order contribution to $\Sigma$: \begin{eqnarray} 
\Sigma^{(2)}(\omega,N) ={ \left ( {\frac{mU_0}{\hbar^2}} \right )}^2 \frac{m}{
(2\pi)^2}\frac{(\hbar \omega_c)^3}{\hbar^2} \sum_{N_1, N_2, N_3}I(N,N_1,N_2,N_3)
\times \nonumber \\ \times\frac{\nu(N_1)[1 -\nu(N_2)][1-\nu(N_3)] +[1 -\nu(N_1)]
\nu(N_2) \nu(N_3)}{\omega +\epsilon(N_1)-\epsilon(N_2)-\epsilon(N_3)} \label{b3} 
\; , \end{eqnarray} where $\nu(N_i)$ is the filling factor of the LL with number 
$N_i$. We use the short-cut notations $N_i$ for $N^\uparrow_i$ and $N^\downarrow_i$ 
in Eqs. (\ref{b3}), (\ref{b4}), and (\ref{b5}). The value $I(N,N_1,N_2,N_3)$ is 
given by  \begin{eqnarray} I(N,N_1,N_2,N_3)=\int_{0}^{\infty}r {\rm d} rJ_0(r\alpha) 
J_0(r\alpha_1) J_0(r\alpha_2) J_0(r\alpha_3) \label{b4} \; , \end{eqnarray} where 
$\alpha_i=\sqrt{2(N_i+1/2)}/l_H$. As in the 3D case \cite{La-Li}, the term 
$\nu(N_1) [1-\nu(N_2)][1-\nu(N_3)]$ in (\ref{b3}) accounts for a process where the
interaction causes the initial particle to jump from the LL $N>N_0$ to the 
LL $N_2$ above the Fermi surface, $N_2>N_0$, so that a particle below the 
Fermi surface in the LL $N_1<N_0$, moves above the Fermi surface to the LL 
$N_3>N_0$ (Figs. 2a, 2b). The term $[1-\nu(N_1)] \nu(N_2) \nu(N_3)$ accounts
for the  analogous process when there is a hole in the initial state (Figs. 2c, 2d).
In a first step, we choose all particles and holes in intermediate states such that 
there is no particle in the LL $N_i$, $N_i \leq N^\uparrow_0$, and no hole in the LL 
$N_k$, $N_k > N^\uparrow_0$, in any intermediate state. The processes with a 
particle involved in the LL $N^\uparrow_0$ will be discussed later.

The way we arrive at Eqs. (\ref{b3}) and (\ref{b4}), rather qualitative, 
could be easily reformulated rigorously in the second order of perturbation theory. 
In such an analysis, the Bessel functions would naturally appear in Eq. (\ref{b4}) as
this function represents the Green function in the coordinate space in the
quasiclassical limit \cite{B-G}. However, such an analysis becomes rather 
sophisticated for higher-order diagrams whereas our simple approach will be 
generalized in Section IV to estimate a diagram of {\it any} order. 

Although the integral (\ref{b4}) can be exactly expressed through a complete 
elliptic integral of the second kind \cite{Manual}, already the simplest 
dimensionality analysis gives an estimate of its value \cite{WE}. This 
is sufficiently accurate for our analysis (only the weak function $\ln N_0$
is missing at $N$, $N_i\approx N_0$): $I \approx l^2_H/N_0$.

The essential property of $\Sigma^{(2)}$ for 2D electrons in a magnetic field 
is that the internal energies are summed in Eq. (\ref{b3}), not integrated as in 
the case of vanishing magnetic field \cite{La-Li}. This is because the spectrum 
of the system is discrete. As a result, the denominator of Eq. (\ref{b3}) contains 
terms with $N+N_1=N_2+N_3$. They describe resonances  when the energy of the 
initial particle (hole) is exactly  equal to the energy of the system in the 
intermediate state, one hole (particle) and two particles (holes) in Fig. 2. One 
distinguishes these resonant processes from non-resonant ones, for which the 
denominator in Eq. (\ref{b3}) differs from that of the  bare Green function by an 
integer of $\hbar \omega_c$. The total $\Sigma^{(2)}$ is given by $\Sigma^{(2)}
=\Sigma^{(2)}_{\rm res}+ \Sigma^{(2)}_{\rm nres}$. Introducing in Eq. (\ref{b3}) a 
new index $j$ instead of $N_1$, $j =N_2+N_3-N-N_1$, one performs the summation over 
indices $N_2$ and $N_3$: \begin{eqnarray} \Sigma^{(2)}(\omega,N)\propto{\left(
\frac{U_0}{l_H^2} \right )}^2 \frac{\hbar \omega_c}{\mu}\sum_{j}\frac{A^2_j}{\omega
-\epsilon(N)-j \hbar \omega_c}\label{b5} \; . \end{eqnarray} The coefficients 
$A^2_j$ contain the filling factors $\nu_i$ from Eq. (\ref{b3}) and some 
combinatorial factors (cf. Eq. (\ref{b6}) and the discussion below). 

The physical effects considered in the present paper are due to resonant terms in
the self-energy part. These terms  have nonzero weight (poles) on the plane $\omega$ 
due to degeneracy of the system, and not the branch cut as for a system with a
continuous spectrum: \begin{eqnarray} {\Sigma^{(2)}}_{\rm res}(\omega, 
N) \propto  \frac{C^2}{\omega -\epsilon(N)} \label{b6} \; . \end{eqnarray} Here
$C^2 \approx k(k-1)V^2A^2_0/2N_0$, where $V$ is given by Eq. (\ref{a7}), and the
combinatorial factor $k(k-1)/2$ accounts for the number of possible
resonances for the LL  with index $N$ such that $k=N-N_0$ for
a particle  and $k=N_0-N+1$ for a hole.

The Green function $G^{(2)}(\omega,N)=1/[\omega - \epsilon(N) - 
\Sigma^{(2)}_{\rm res}(\omega,N)]$, corresponding to $\Sigma^{(2)}_{\rm
res}$, is given by \begin{eqnarray} G^{(2)}(\omega,N)= \frac{1/2}{\omega - 
\epsilon(N) - C} + \frac{1/2}{\omega - \epsilon(N) + C} \label{b7} \; . 
\end{eqnarray} As a result of resonance, the initial LL with energy $\epsilon_0
=\epsilon(N)$  splits up into two sublevels with energies $\epsilon_\pm  
=\epsilon(N)\pm C$ and with the corresponding residues $1/2$. With the use 
of the diagram technique, one is thus able to describe the situation, typical for 
any degenerate system, when an external interaction -- in the present case 
interparticle interactions -- leads to a splitting of the initially degenerate 
LL with a partial lifting of the degeneracy.

So far we neglected the possibility of a particle to move onto the vacant place
in the LL $N^\uparrow_0$. Let us now consider the dependence of $\Sigma^{(2)}$ on 
the filling factors $\nu(N_i)$ in Eqs. (\ref{b3}),(\ref{b6}). Recalling  that 
$\nu(N_i)=1$ for 
$N_i<N^\uparrow_0$, $\nu(N_i)=1 -\nu_0$  for $N_i=N^\uparrow_0$, and $\nu(N_i)=0$ 
for $N_i>N^\uparrow_0$, one concludes that processes not involving a particle in the 
LL $N^\uparrow_0$ (cf. Fig. 3a), can exist for all values of $1-\nu_0$. They are 
symmetric with respect to  spin. The characteristic order of renormalization of the 
energy by these processes is $C_{l} \sim V/N^{1/2}_0$. We shall call them ``large'' 
processes. Processes in which a particle is created in the LL $N^\uparrow_0$, 
contain an additional small parameter in the numerator of Eq. (\ref{b6}). For them, 
the renormalization of energy  is of the order $C_{s}\sim V[(1-\nu_0)/N_0]^{1/2}$. 
Such ``small'' processes occur only when $\nu_0$ is different from 1: for $\nu_0=1$ 
only ``large'' processes survive. Examples of ``small'' and ``large'' processes 
for the LL $N^\downarrow_0+2$ are given in Fig. 3. The ``small'' processes
are asymmetric with respect to the spin of the initial particle:
Electrons or holes in a LL with spin  down, $N^\downarrow_i$, can
directly interact with the electrons and holes in the LL $N^\uparrow_0$,
giving rise to processes  where {\it only} these two LL's, $N^\downarrow_i$ and
$N^\uparrow_0$, are involved (cf. Fig. 3b). For a LL with spin up, this
is forbidden because of the interaction chosen. 

The poles of the terms in $\Sigma^{(2)}_{\rm res}$, for a 3D system without 
\cite{Gal} and with \cite{P-P} magnetic field and a 2D system without magnetic 
field  \cite{Blood}, determine the damping of quasiparticles $\gamma \sim (m 
U_0/ \hbar^2)^2 \, (p-p_f)^2/m$. Using $p_f(p-p_f)/m=\hbar \omega_c$, this 
result is also recovered from Eq. (\ref{b3}) in the limit $H \rightarrow 0$. 
However, when $H \neq0$, the square root is taken from $V^2/N_0$ due to the 
resonance. That is, the renormalization of the spectrum described by the terms
$\Sigma^{(2)}_{\rm res}$ is proportional to the {\it first} power of the parameter 
of perturbation theory (cf. (\ref{b7})), in contrast to the usual situation. 
Analogously, instead of a factor $(p-p_f)^2/m$, characterizing the damping of 
quasiparticle with momentum $p$ remote from the Fermi surface, the splitting of the 
LL $N_0\pm k$ is proportional to the {\it first} power of analogous parameter $k$. 
The resonant terms do not renormalize the total filling factor of each LL. 
Similarly to the traditional theory of the Fermi liquid \cite{La-Li}, it is 
the non-resonant processes $\Sigma^{(2)}_{\rm nres}$ which renormalize the 
filling factors of LL's \cite{WE}. The corrections to the bare values are of 
second order in the perturbation parameter.

Thus, already in second order of perturbation theory, one finds that the 
renormalization of our system should be essentially different from that of the known 
systems with a dispersion in the spectrum. In the next sections, we shall consider 
the higher orders of perturbation theory, concentrating  on the contributions to 
$\Sigma$ which determine the first-order corrections to the spectrum.

\section{Rules for estimates}

Having understood the character of renormalization of the spectrum in second-order 
perturbation theory -- its splitting with partial lifting of degeneracy -- 
we proceed to higher orders of perturbation theory. In this section we formulate the 
general rules, which hold for any LL with $N_i \gg 1$, according to which the most 
essential diagrams should be selected. 

In calculating a given diagram $\Sigma^{(n)}$, one has to sum over all 
independent frequencies and integrate over all independent momenta \cite{AGD}. 
These two procedures are independent of one another and will be shown to produce a
small parameter $1-\nu_0$ or $1/N_0$.

Indeed, as a result of the summation, there will appear a product of different 
filling factors $\nu_{i}$ and $1-\nu_{j}$, so that $\nu_{i}$ ($1-\nu_{j}$) 
corresponds to a propagator of a hole in the  $i$th LL (an electron in the 
$j$th LL) \cite{LUT}. From all possible processes, one should select only
those with the minimal power of $1-\nu_0$. In general, the renormalized
values of the filling factors should be taken. However, we are searching
for the correction to the spectrum of the first order in the parameter
$V/\hbar \omega_c$. Since the corrections to the filling factors are of
second order, it is permissible to use the bare filling factors.

The parameter of perturbation theory $V/\hbar \omega_c$ was introduced in Eq. 
(\ref{a7}) of Section II. The $n$th power of $V$, having the dimensionality of 
energy, enters each diagram of $n$th order. To preserve the proper 
dimensionality of this correction to the self-energy  $\Sigma^{(n)}$, the 
denominator should have the dimensionality of the $(n-1)$th power of energy. 
Applying the Luttinger algorithm for the determination of the self-energy part 
\cite{LUT}, one finds \begin{eqnarray} \Sigma^{(n)}(\omega, N) \propto \frac{V^n} 
{T^p \,\Pi_i(\omega-k_i \hbar \omega_c -\epsilon_N)^{s_i}} \label{c1}  \; , 
\end{eqnarray} with integers $k_i$, $s_i$,  $p$ such that $\sum_i s_i+p =n-1$.
The second-order contribution, Eq. (\ref{b6}), also has the form of Eq. (\ref{c1}). 
According to Ref. \cite{LUT}, each cut of a diagram contributes a factor ($\sum_i 
\epsilon_i$) to the denominator in Eq. (\ref{c1}), where $\epsilon_i$ is the energy 
of an electron in this cut (for a hole $-\epsilon_i$ should be taken). A cut 
containing one external frequency $\omega$ contributes a power of $(\omega-k_i \hbar 
\omega_c -\epsilon_N)$, where the integer $k$ denotes the difference from the bare 
energy $(\omega-\epsilon_N)$. These factors can be identical for different cuts; 
this is accounted for by integers $s_i$. In principle, powers of $T$ could also 
occur in the denominator of Eq. (\ref{c1}) from the third order of perturbation 
theory on. A power of $T$ could occur each time two propagators (the incoming and 
the outgoing) with the external frequency $\omega$ are present in a cut of a 
diagram. So, one could expect that the cut A-B  of the diagram shown in Fig. 4 
gives rise to the first power of $1/T$ in Eq. (\ref{c1}) on condition that 
$\epsilon^\downarrow_{1}+\epsilon^\uparrow_{2}= \epsilon^\downarrow_{3}+
\epsilon^\uparrow_{4}$. Such terms would give diverging contributions to
$\Sigma$ at $T \rightarrow 0$. However, powers $1/T^p$ with $p>0$ do not arise 
in {\it any} order of perturbation theory for our contact interaction, provided 
that the bare filling  factors are used. This is illustrated most easily by the 
example of the diagram shown in Fig. 4. The propogators in the cut A-B, giving 
rise to a term $T$ in the denominator, at the same time introduce the product 
of filling factors $\nu^\uparrow_{1} \nu^\downarrow_{2} (1-\nu^\downarrow_{3}) 
(1-\nu^\uparrow_{4})$ in the numerator. It is apparent that for each choice of 
intermediate indices $N_i$, leading to a resonance, at least one of the four 
factors in the numerator must be zero. The case when all intermediate 
particles $i=1, ...  4$ are in the LL $N^\uparrow_0$ is not possible, because 
particles with the same spin do not interact.

It follows from the aforesaid that resonant processes, resulting in the 
maximally possible $(n-1)$th power of the singular denominator in Eq. 
(\ref{c1}) with $p=0$; $s_i=n-1$, $k_i=0$; $s_j=0$, $k_j \neq 0$, determine 
the spectrum near the bare LL energy $m\hbar \omega_c$ in the $n$th order of 
perturbation theory. With accuracy $(V/\hbar\omega_c)^2$, the summation over 
the LL's in Eq. (\ref{a6}) is no longer important. In all the intermediate 
Green functions we can keep only the LL's, leading to a $\Sigma(\omega)$ which 
is resonant with the bare Green function $G_0(\omega)=1/(\omega-m\hbar\omega_c) 
=1/x$. Here, the notation \begin{eqnarray} x=\omega-m\hbar\omega_c \label{c2}  
\end{eqnarray} has been used to denote the correction to the spectrum. Since the 
full Green function for the $m$th LL is $G(x)=1/[x-\Sigma(x)]$, the spectrum of 
interacting system is determined by the solution of the equation $x=\Sigma(x)$ 
with resonant $\Sigma$.

Tracing back the origin of the dimensionless factor $\hbar \omega_c/\mu =1/N_0$ 
in Eq. (\ref{b5}), one is led to the conclusion that it is due to the  
momentum-conserving $\delta$-function in Eq. (\ref{b2}). Analogously, each 
independent $\delta$-function in a vertex of any diagram will contribute a factor 
$1/N_0$ to the estimate. Further we will mark each independent law of conservation 
of momentum by a circle around the vertex. 

These rules will allow us to obtain quickly the dependence of any
diagram on the values $1/N_0$, $1-\nu_0$, $V$, and $x$.

\section{Higher-order processes} 

In the present section the essential higher-order processes are introduced
and estimated. Depending on the index of the LL, different processes are important.

The classification of ``large'' and ``small'' processes, given in Section
III, holds also for  higher orders of perturbation theory.  For LL's $N_0$ and 
$N_0+1$, only ``small'' processes are possible, whereas for LL's, situated further 
away from $N_0$, ``small'' and ``large'' processes, as well as their combinations 
occur. Examples can be understood by looking at Fig. 3. The only exception is the LL 
$N^\uparrow_0-1$, for which only  ``large'' processes exist (cf. discussion below).

Since ``small'' processes contain additional factor $1-\nu_0 \ll 1$, it is clear 
that the renormalization is governed by ``large'' processes, if any (the exact 
formulation is given in Section VII). Among all ``large'' processes, one should 
select those with the minimal power of 
the small parameter $1/N_0$, i.e. with the minimal number of independent 
$\delta$-functions in a given order of perturbation theory. Such decay processes 
may be interpreted as a generalization of a second-order one, in which one of the 
intermediate particles or holes in turn undergoes a $\Sigma^{(2)}$ process, etc.
This is shown in Fig. 5. An essential feature of this process is the necessity 
for a particle or hole to decay into particles and holes situated nearer to 
the LL $N^\uparrow_0$  in each following process. This means that ``large'' 
processes up to $k-1$th ($k$th) order of perturbation theory are possible for 
the LL $N_0+k$ ($N_0-k$). The estimate of the decay process of the $2n$th order, 
carried out with the rules of Section IV, gives \begin{eqnarray} \Sigma^{
(2n)}_{\rm dec,\, l}(x)\propto \frac{V^{2n}}{N^n_0 x^{2n-1}} \label{d1} \; . 
\end{eqnarray} Any ``large'' decay processes  result in the correction to the 
spectrum of the same order $x \sim V/\sqrt{N_0}$ in each order of perturbation 
theory.

Analogously to the conventional Fermi-liquid theory \cite{La-Li}, it follows from
energy conservation that all the intermediate particles and holes must lie between 
$N_0-k$ and $N_0+k$ in any diagram for the LL  $N_0\pm k$. Hence, the 
effective energies of the LL's further away from the LL $N_0$ cannot be found 
without knowledge of the energies of nearer LL's. Let us select the essential 
diagrams for the LL's near $N_0$.

1. For the LL $N^\uparrow_0$, neither an electron- nor a hole-like
excitation  can interact with the electrons and holes of this LL,
because the spin projection is the same.

2. For the LL's $N^\downarrow_0$ and $N^\downarrow_0+1$, there are two series 
of diagrams, containing the minimal power of one of our two small parameters, 
$1/N_0$ and $1-\nu_0$.  The minimal power of $1/N_0$ is contained in the 
``small'' decay processes. In contrast to ``large'' decay processes for LL's 
further away from $N_0$, the ``small'' ones exist in any order of perturbation theory. 
With including the next $\Sigma^{(2)}$ in the self-energy, an additional particle  
is created in the unoccupied part of the LL $N^\uparrow_0$. This contributes  
each time a factor $1-\nu_0$ to the estimate. For the ``small'' decay  process 
of $n$th order one has  \begin{eqnarray} \Sigma^{(2n)}_{\rm dec, \, s}(x)\propto 
\frac{(1-\nu_0)^n V^{2n}}{x^{2n-1}N^n_0} \label{d2} \; . \end{eqnarray} The correction 
to the energy is of the same order for all these processes $x \sim V (1
-\nu_0)^{1/2}/N^{1/2}_0$.

The ladder graphs represent the second essential series. An example of such a graph 
is given in Fig 6a. The estimate of a ladder graph, where the interaction of an 
initial hole in the LL $N^\downarrow_0$ and a particle in the $N^\downarrow_0+1$ 
with holes in the LL $N^\uparrow_0$ is accounted for (Figs. 6b, 6c, respectively), 
contains only the first power of the parameter $1-\nu_0$ in any order of perturbation 
theory (because only one electron is created in the LL $N^\uparrow_0$ in intermediate 
states), but higher powers of $1/N_0$ in comparison with the previous series:  
\begin{eqnarray}  \Sigma^{(2n)}_{\rm lad}(x) \propto 
\frac{(1-\nu_0)V^{2n}}{x^{2n-1}N^{2n-1}_0} \label{d3} \; . \end{eqnarray}

3. For the LL's $N^\uparrow_0+1$ and $N^\uparrow_0-1$, only the processes of second 
order are essential, the ``small'' one for $N^\uparrow_0+1$ and the ``large'' one 
for $N^\uparrow_0-1$, presented in Figs. 7a, 7b. The ladder graphs are not 
important for the LL $N^\uparrow_0+1$: Both possibilities -- the interaction of the 
intermediate particle in the level $N^\downarrow_0+1$ with electrons in the LL 
$N^\uparrow_0$ and the interaction of the intermediate particle in the LL 
$N^\uparrow_0$ with the holes in the LL $N^\downarrow_0$ -- give a power of the 
factor  $(1-\nu_0)^k$ with $k>1$ in any order of perturbation theory, beginning from 
the third. This is because more than one electron is created in the LL $N^\uparrow_0$. 
As for decays, the possibilities for the particle in the LL 
$N^\downarrow_0+1$ and the hole in the LL $N^\downarrow_0$ to decay are already 
accounted for by their renormalized Green functions. The hole in the LL 
$N^\uparrow_0-1$ does not feel the deviation of the filling factor $\nu_0$ from 
unity directly, because it does not interact with the electrons and holes in the LL 
$N^\uparrow_0$, i.e. there exist only ``large'' processes for this LL.

\section{Renormalization of the LL's near the LL $N_0$}

In this section we carry out the program discussed above: for each LL we select
the series  $\Sigma^{(n)}(x)$ and sum them up, using the estimates of the 
previous section. Thus, we derive the Dyson equations (DE's)  and solve them. 

1. In the LL $N^\uparrow_0$, neither a particle nor a hole can interact. This 
means that the full Green function coincides with the bare one and the 
energy does not change. 

2. For the LL's  $N^\downarrow_0$ and $N^\downarrow_0+1$, we have introduced 
two infinite series of relevant diagrams, which have minimal powers
of one of our two small parameters, $1/N_0$ or $1-\nu_0$. These values 
can be varied independently, so it is possible to study two limiting situations, 
$\xi \ll 1$ and $\xi \gg 1$, where we have defined a dimensionless parameter
\begin{eqnarray} \xi=N_0(1-\nu_0) \label{e1} \; . \end{eqnarray}

If $\xi\ll 1$, it follows from the comparison of  Eqs. (\ref{d2}) and (\ref{d3}) 
that the ladder graphs should be selected. The resulting DE is shown in Figs. 8a, 8b.

If $\xi\gg 1$, the decay processes become more important. However, it would be 
wrong to restrict oneself to the decay processes only. If this were done, then 
the DE would have the form of  Figs. 9a, 9b with the bare vertex. After 
providing summation over frequencies $\omega_1$ and $\omega_2$ one would 
obtain \begin{eqnarray} \Sigma(x)=\frac{(1-\nu_0)V^2}{[x-\Sigma(x)]N_0} \label{e2}
\; , \end{eqnarray} leading to a physically meaningless result, viz. that 
the full Green function, corresponding to Eq. (\ref{e2}), would have no poles. To 
select the correct series, and obtain the spectrum of the interacting system, 
let us notice that Eq. (\ref{e2}) is singular under the condition $x=\Sigma(x)$, 
determining the correction to the spectrum. Therefore, it is necessary to take 
into account terms with higher powers of the difference $x-\Sigma(x)$ in the 
denominator. One checks that all essential diagrams are included after the 
bare vertex is replaced with the vertex shown in Figs. 9a and 9b.

It is worth noticing that the Green function with the self-energy part Eq. (\ref{e2})
is mathematically well defined. Though it has no poles, its analytical
properties are determined by the cut between two branching points, $x=\pm 2V
[(1-\nu_0)/N_0]^{1/2}$, on the complex plane $x$. However, a careful diagrammatic 
analysis reveals an effective dressing. (Its physical significance is discussed below.)
As the result, the Green function will be shown to acquire poles.

Comparison of Figs. 8 and 9 shows that the bare Green functions 
are used in the DE of Fig. 8, instead of the full ones of Fig. 9. 
This implies that the solution of a DE for the case $\xi \ll 1$ 
should be recovered after omitting the self-energy part in a DE for the case 
$\xi \gg 1$. Further on, we treat DE for both cases simultaneously. 

The dependencies on the momenta in the DE of Figs. 8a, 9a, are essentially
simplified by using the Fourier transform \cite{B}. Accordingly, we search
for the full vertex of the LL $N^\downarrow_0$ in the form \begin{eqnarray}
\nonumber \Gamma(p_1-p_3,p_1-p_4)=\int {\rm d} q_x \, \exp[iq_x(p_1-p_3)] \, 
\Gamma(q_x,p_1-p_4) \; . \end{eqnarray} 
The vertex of the DE gives \begin{eqnarray} \Gamma(q_x,p_1-p_3;\omega_1+
\omega_2)=W_{N_0,N_0}(q_x,p_1-p_3) \sum_{m}^{2N_0} E_m \times \nonumber\\ 
\frac{i(\omega_1 +\omega_2) -\epsilon_{0\uparrow}- \epsilon_{0\downarrow} 
-[\Sigma]} {i(\omega_1 +\omega_2) -\epsilon_{0\uparrow}- 
\epsilon_{0\downarrow} +E_m-[\Sigma]} \label{e3} \; . \end{eqnarray}
Here, apart from the notations introduced in Fig. 8a, 9a, 
\begin{eqnarray} W{n_1,n_2}({\bf q})= U_0 \exp(-{\bf q}^2/2)
L_{n_1}({\bf q}^2/2)L_{n_2}({\bf q}^2/2) \label{e4} \; , \end{eqnarray} 
where $L_n$ is the Laguerre polynomial;
$\epsilon_{0\uparrow}$ ($\epsilon_{0\downarrow}$) is the bare energy of
the LL $N^\uparrow_0$  ($N^\downarrow_0$), $m$ is even, and 
\begin{eqnarray} E_m=V\int {\rm d} q^2 \exp(-q^2)[L_{N_0}(q^2/2)]^2L_m(q^2) 
 \label{e5} \; . \end{eqnarray} The term $[\Sigma]$ in Eq. (\ref{e3}), as well as 
in the following equations, should be omitted for the case $\xi\ll 1$ and kept for 
$\xi\gg 1$. An estimate shows that $E_m \sim V/N_0$ for $1 \ll m \ll 2N_0$, which 
contain the range of $m$ essential for summation. Using values $E_m$, it is not 
difficult to rederive the estimates (\ref{a7}) and (\ref{d3}).

To calculate the self-energy, given by \begin{eqnarray} \Sigma(\omega)=T^2\, 
\sum_{\omega_1 \omega_2} \, G(\omega_1)G_0(\omega_2)G_0(\omega+\omega_1
+\omega_2) \Gamma(\omega+\omega_2)  \label{e6} \; ,  \end{eqnarray} 
some assumptions about the full propagator $G$ of the LL $N^\downarrow_0$ have to be
done. Later we will see that a propagator with several poles, one of them 
having the prevailing spectral weight, satisfy the requirement
of selfconsistency. The following calculations will determine the values of this 
shift and explain the meaning of the poles.

Performing first the summation over the frequency $\omega_1$, one obtains
\begin{eqnarray}  \Sigma(\omega) \propto  T\sum_{\omega_2,\,m}
\frac{E^2_mG_0(\omega_2)}{i(\omega_2+\omega)- \epsilon_{0\uparrow}
-\epsilon_{0\downarrow} +E_m- [\Sigma]}  \label{e7} \; .  \end{eqnarray}
The final result of the summation over $\omega_2$ and integration over the
independent momenta reads: \begin{eqnarray} \Sigma(x) = \sum_m
\frac{E^2_m(1-\nu_0)}{x+E_m-[\Sigma]}  \label{e8} \; . \end{eqnarray}
Equation (\ref{e8}) determines $2N_0+1$ poles of the Green function, $2N_0$ of them
lie near $x_m \sim [\Sigma] -E_m$. Considering the case $\xi\ll 1$ and assuming $x 
\ll E_m$, one obtains $\Sigma \approx (1-\nu_0)V$. The ($2N_0+1$)th pole is
obtained with use of $x=\Sigma$, \begin{eqnarray}x \approx (1-\nu_0)V  \label{e9}
\; . \end{eqnarray} One  checks that $x$ is indeed less than $E_m$, $(1-\nu_0)V \ll 
E_m$. For the case $\xi\gg 1$ one obtains the same result after substituting the 
condition $x=\Sigma(x)$, determining the correction to the spectrum, into Eq. 
(\ref{e8}). Representing the full Green function as a sum of singular terms, one 
proves that the total weight of first $2N_0$ poles is indeed smaller than that of 
the ($2N_0+1$)th by factor of $1-\nu_0$, which justifies our initial assumption. 

Appearance of $E_m$ in the denominator of Eq. (\ref{e8}) is not accidental. These values
are nothing but the eigenvalues of energy of a system of two interacting holes. They 
are discrete as they are classified by eigenvalues of total angular momentum
\cite{Byby}. Thus, one interpretates the considered effect as a dressing of the bare
propagator by a ``collective mode" -- two interacting holes. As a result of this, the
bare energy of the LL $N^\downarrow_0$ is shifted and $2N_0$ poles with small
residues appear, reminiscent of the eigenvalues of the collective mode. Further we 
will consider only the shifted level. 

For the analysis of the DE for the LL $N^\downarrow_0+1$, it is most natural 
to make use of the representation of the total transverse momentum \cite{Braz}, 
which simplifies the dependence of the vertex on the momenta: \begin{eqnarray}
\Gamma(p_1-p_4,q_y; \omega_1-\omega_4)=F_{N_0,N_0+1}(p_1-p_4,q_y) \times
\nonumber\\ \frac{(i(\omega_1-\omega_4)+ \epsilon_0 -\epsilon_{N_0+1} 
-[\Sigma] )}{i(\omega_1 -\omega_4)+ \epsilon_0 -\epsilon_{N_0+1}
-[\Sigma] -F_{N_0,N_0+1}(p_1-p_4,q_y)}  \label{e10} \; . \end{eqnarray}
In Eq. (\ref{e10}), $F_{N_0,N_0+1}$ is the Fourier transform of the function 
$W_{N_0,N_0+1}(x, y)$, performed on both of the coordinates $x$ and $y$, and  
$\epsilon_0$ ($\epsilon_{N_0+1}$)  denotes the energy of the LL 
$N^\uparrow_0$ ($N^\downarrow_0+1$). As in the  previous case, we 
have \begin{eqnarray} \Sigma = (1-\nu_0)\int {\rm d} p {\rm d} q \, 
\frac{[F_{N_0,N_0+1}(p,q)]^2}{x-[\Sigma]+F_{N_0,N_0+1}(p,q)}  \label{e11}
\; , \end{eqnarray} which leads to the same results as Eq. (\ref{e8}).

The bare propagator for the LL $N^\downarrow_0+1$ is dressed by another ``collective 
mode", interacting electron and hole. The eigenvalues of energy for such a system 
are classified by a continues parameter, the total momentum \cite{BIE,KH}.
That is way the integration is present in Eq. (\ref{e11}). 

Thus, the bare energies of the LL's $N^\downarrow_0$ and $N^\downarrow_0+1$ are 
simply shifted in both regimes. What is more, the values and the signs of the shift 
are the same. In order to prove this, we multiply both sides of the expansion 
\cite{Manual} $2V[L_n(r)]^2=\sum_k^n L_{2k}(2r) E_k$ on $e^{-r}$, then integrate 
over the variable $r$, and obtain $W_{N_0,N_0+1}(0)=\sum_k^{2N_0} E_k$. This means 
that $\Sigma$ from Eqs. (\ref{e8}) and (\ref{e11}) are identical on condition  
$x=\Sigma$.

3. The  LL's $N^\uparrow_0+1$ and  $N^\uparrow_0-1$ were shown in Section V 
to be renormalized by the second-order processes. Taking into account the 
identity of the shifts for LL's $N^\downarrow_0+1$ and $N^\downarrow_0$, this 
results in the splitting of the  LL's $N^\uparrow_0-1$ and $N^\uparrow_0+1$ 
in two sublevels for both $\xi \gg 1$ and $\xi \ll 1$ in the manner, similar 
to Section III. The splitting of the LL $N^\uparrow_0+1$ 
has the order of $V(1-\nu_0)^{1/2}N^{1/2}_0$, and that of
the LL $N^\uparrow_0-1$ is of order $VN^{1/2}_0$.

\section{Renormalization of the LL's further away from the LL  $N_0$}

Once the energies of the LL's near $N_0$ have been established, one can proceed to  
LL's, situated further. Here, the renormalization is determined by another
dimensionless parameter than in Section VI because of the coexistence of ``small'' 
processes, $x\sim x_1=V(1-\nu_0)$, ``large'' processes, $x\sim x_2=V/ \sqrt{N_0}$, 
as well as their combinations. The spectrum essentially differs in the following 
cases: (i) $\sqrt{N_0}(1-\nu_0) \ll 1$; (ii) $\sqrt{N_0}(1-\nu_0) \approx 1$;
(iii) $1/ \sqrt{N_0} \ll  1-\nu_0 \ll 1$. In the present paper only the cases (i) 
and (iii) are analyzed. In case (ii) ``large'', ``small'' processes, and their 
combinations are equally important and we were not able to select the 
essential series.

(i) For $\sqrt{N_0}(1-\nu_0) \ll 1$, the renormalization is governed by ``large'' 
decay processes, symmetric with respect to spin. Schematically one writes 
\begin{eqnarray} \Sigma(x) \propto \Sigma_{\rm dec}(x)+\frac{V^3}{x^2 N^2_0} 
+\frac{V^2(1-\nu_0)}{x N_0}  \label{f1} \; , \end{eqnarray} where the first term 
represents contributions from the ``large'' decays,  the second and the third terms 
estimate the third-order ``large'' ladder-graph and the second-order ``small'' 
process, respectively. The second-order processes  for the LL $N^\downarrow_0+2$ are 
shown in Fig. 3. The characteristic order of the renormalization $x \sim V/\sqrt{N_0}$ 
is given by the first term in Eq. (\ref{f1}). One proves that any correction to  
$\Sigma$   from ``small'' processes is smaller than one from ``large'' non-decay 
processes: The last two terms in Eq. (\ref{f1}) are of the same order only when $1-
\nu_0 \approx 1/\sqrt{N_0}$, which contradicts to $\sqrt{N_0}(1
-\nu_0)\ll 1$. Among all ``large'' processes, the decays should be selected.  

According to Section V, there are $k$ possible ``large'' decay processes for the LL's
$N_0+k+1$ and $N_0-k$. Since the intermediate particles and holes in any ``large''
process, occurring for the LL's $N_0+k$ and $N_0-k$, must lie in the LL's, situated
closer to the LL $N_0$, $N_0\pm |k-1|$, the Green function of each following LL
depends on the propagators of all the previous LL's. One determines the
renormalized spectrum recursively, gradually 
increasing index $k$. 
The LL's $N_0$ and $N_0+1$ are not renormalized within the accuracy $x \sim 
V/\sqrt{N_0}$. The next two levels, $N_0-1$ and $N_0+2$, are split into two 
sublevels by the second-order process in the manner described in Section III.

The fourth-order decay process for the LL's $N_0+3$ and   $N_0-2$  can be
considered as a second-order process, in which the intermediate decay of the 
particle in the LL $N_0+2$ (or hole in the $N_0-1$) is described by its 
Green function Eq. (\ref{b7}). Both processes give the same result: \begin{eqnarray} 
\Sigma^{(4)}_{\rm dec}(x)=\frac{C^2_2\,V^2}{2N_0\,(x-C_1V/\sqrt{N_0})} 
+\frac{C^2_2\,V^2}{2N_0\,(x+C_1V/\sqrt{N_0})} \label{f2} \; , \end{eqnarray}
where the constants $C_1$, $C_2$ are of order of unity. The Green function 
corresponding to Eq. (\ref{f2}), has the form \begin{eqnarray} \nonumber G^{(4)}
(x) = \frac{1}{x\,[1 - C^2_2 V^2/(N_0 x^2 -C^2_1 V^2)]} = \frac{\gamma_1}{x}
+\frac{\gamma_2}{x-V a/\sqrt{N_0}}+ \frac{\gamma_2}{x+V a/\sqrt{N_0}} 
\end{eqnarray} with constants $a=\sqrt{C^2_1+C^2_2}$, $\gamma_1=C^2_1/2a^2$, 
and $\gamma_2=C^2_2/2a^2$. We again observe a splitting of the bare LL's, 
$x=0$  and $x=\pm Va/ \sqrt{N_0}$ such that the total quasiparticle weight
does not change, $\gamma_1+2\gamma_2=1$.

The energies of higher LL's can be found analogously. Like the LL's considered 
above, they are also split into sublevels with characteristic $x \sim  x_2=
V/\sqrt{N_0}$, and the total quasiparticle weight of each LL is equal to its bare 
value. It can be seen that the number of sublevels growths rapidly when moving 
away from the LL $N_0$, however we were not able to derive any universal dependence. 
The resulting spectrum in this regime is presented  in Fig. 10.

The typical value of the splitting $x_k$ grows according to Eq. (\ref{b6}), $x_k \sim 
kV/\sqrt{N_0}$, as the index of the LL $N_0\pm k$ deviates from $N_0$. Thus, one  
can formulate the analog of the Landau criterion: The Fermi-liquid description 
is applicable to our system up to the LL $N_0 \pm k$, where $k$ is such that 
two neighbouring  LL's do not overlap:  \begin{eqnarray} kV/ \sqrt{N_0} \ll 
\hbar \omega_c  \label{f3} \; . \end{eqnarray}

(ii) Let us proceed to the next regime of renormalization, $1/ \sqrt{N_0} \ll
1-\nu_0 \ll 1$. Now, the correction to the energy of any LL due to renormalization
of the lower LL's $N^\downarrow_0$ and $N^\downarrow_0+1$, 
found in the case $\xi \gg 1$ to be  $x \sim x_1=V(1-\nu_0)$, exceeds that of the 
``large'' decay processes, $x\sim x_2= V/\sqrt{N_0}$.

The asymmetry we have found near the LL $N_0$, when the energy $x^\uparrow$
of the LL's $N^\uparrow_0-1$ and $N^\uparrow_0+1$ with spin up and the energy 
$x^\downarrow$ of the LL's $N^\downarrow_0$ and $N^\downarrow_0+1$ with spin 
down have different orders of magnitude, $x^\uparrow 
\ll x^\downarrow$, is preserved also for higher numbers of the 
LL's. Thus, the renormalization appears to depend on the spin of a LL.

For a particle placed in a LL with spin down, say $N^\downarrow_0+2$ (Fig. 3), 
there exist ``large'' processes. However, their self-energy part is not resonant 
with the bare energy anymore due to different corrections to energies
of the lower LL's. The estimate of the ``large'' second-order process, 
\begin{eqnarray} \Sigma^{(2)}_\downarrow(x) \propto V^2/[N_0(x^\downarrow+
x^\uparrow_{N_0}- x^\downarrow_{N_0+1}-x^\uparrow_{N_0+1}]  \label{f4} \; , 
\end{eqnarray}  with $x^{\uparrow, \, \downarrow}_j$ denoting the  
energy of the LL $N^{\uparrow, \, \downarrow}_j$, gives the shift of the energy 
$x^\downarrow \sim  V(1-\nu_0)$.

For a LL with spin up, this ``detuning" does not take place. As an example, let us 
consider LL $N^\uparrow_0+2$. The estimate of the second-order process, analogous to 
(\ref{f4}), \begin{eqnarray} \Sigma^{(2)}_\downarrow(x) \propto V^2/[N_0(x^\uparrow+ 
x^\downarrow_{N_0} - x^\downarrow_{N_0+1}-x^\uparrow_{N_0+1}]  \label{f5} \; , 
\end{eqnarray} yields the correction of energy  $ x^\uparrow \sim V/\sqrt{N_0}
\ll V(1-\nu_0)$. 

It is not difficult to generalize Eqs. (\ref{f4}) and (\ref{f5}) to understand how the
renormalization of the  LL's $N^\downarrow_0$ and $N^\downarrow_0+1$ affects the
other LL's. For a LL with spin down, the initial excitation jumps in a way that among 
the intermediate states only one state with spin down and two states with spin up are 
present. This leads to a shift of the level by an amount of $x^\downarrow \sim  V(1-
\nu_0)$. For a LL with spin up, one can always choose a hole and an electron in the 
intermediate states in the LL's with spin down, so that the initial position of the LL 
does not change.

Thus, within accuracy $x \sim  V(1-\nu_0)$, one effectively has a renormalization of 
the Zeeman splitting. The resulting spectrum is shown in Fig. 11. Let us notice
that there are LL's with opposite spin projections having the same spacing. As a
result, the LL's are split analogously to the first regime at the smaller 
scale of energy $x \sim V\sqrt{N_0}$.

The Fermi-liquid description is applicable if \begin{eqnarray} V(1-\nu_0) \ll 
g \hbar \omega_c  \label{f7} \; , \end{eqnarray} so that the Zeeman-split LL's 
do not overlap.

\section{Conclusions}

In the present paper, the spectrum of excitations of quasiparticle type is 
studied for the 2D electron gas with contact interactions within the conventional 
perturbative approach. The case of a weak magnetic field is considered, i.e. 
many Landau levels are filled. 
It is assumed that the upper Landau level $N_0 \gg 1$ is almost
completely filled, $1-\nu_0 \ll 1$.

For our model, an effect characteristic for a degenerate system, viz. a splitting
of levels by virtue of interactions, is found and described. Due to absence of 
dispersion and due to the equal spacing between the Landau levels, resonances occur
which {\it split} the bare spectrum. The resonances will appear for any kind of 
interactions, however, we were able to solve the problem only for contact interactions.

The spectrum of the interacting system has been found in two regimes. In the
first regime, where $N^{1/2}_0(1-\nu_0) \ll 1$, the bare LL's are split. A scheme is 
suggested making it possible, in principle, to obtain the number of sublevels, 
their energies and filling factors. In the second regime, where $1/N^{1/2}_0 \ll 
1-\nu_0 \ll  1$, the splitting is accompanied by an effective shift of the 
bare Zeeman-split LL's. 
In the case intermediate between these two regimes, one is not able to select the 
essential processes. The spectrum found allows us to derive the criteria (\ref{f3}) and 
(\ref{f7}) determining the applicability of the Fermi-liquid description. 

To our knowledge, the splitting of the LL's has not been discussed
in the literature yet; the Zeeman-splitting is known to be affected by long-range 
interactions  for strong \cite{Ando} and intermediate \cite{MAC} magnetic fields. 
Apart from the strength of the field, another essential difference between Refs.
\cite{Ando,MAC} and the second regime is that for  long-range interactions
the enhancement of the $g$-factor occurs already in the first order of perturbation, 
whereas for the present model higher-order corrections are also important. However, 
after accounting for all the essential processes, we obtain the similar final result: 
a shift of the spin-split LL on the value  $x \sim V(1-\nu_0)$. The electron gas in 
a weak magnetic field and interacting via Coulomb potential was studied in Ref.
\cite{A-G}. There, the renormalization of the LL $N^\uparrow_0$ was governed by two 
characteristic energy scales: the first determined the energy gap in the tunneling 
density of states at a LL and the second described the characteristic shift of the 
Zeeman-split LL. In Refs. \cite{Ando,MAC,A-G}, a completely different interaction 
was used: the effective interaction between electrons in the LL with the same index 
and the spin projection was important, whereas in our model only electrons with 
different spin interact. Keeping in mind the similarities and differences in spectra 
obtain in Refs. \cite{Ando,MAC,A-G}, it would be interesting to find out by which way 
the spectrum changes with change of interactions and strength of the magnetic field.

In the present approach, the essential results arise due to energy 
denominators in the self-energy part $\Sigma$. They can be calculated exactly 
for any diagram. Although some approximations were used while estimating the 
dependence of any given diagram on the momenta, the character of renormalization
is found {\it exactly}. For instance, to predict the same value of shift for  
the LL's $N^\downarrow_0$ and $N^\downarrow_0+1$ no approximation is needed, 
as soon as the essential series of diagrams is established.

We restricted ourselves to accounting for the physical effects arising in 
the minimal order of the small parameters involved.  In this approach, only 
the bare energies of the LL's are renormalized, but not their filling factors:  
The corrections to spectrum are of first order in the parameter $V/\hbar \omega_c$ 
and the bare filling factors are renormalized in second order of perturbation 
theory.  One can hardly predict what kind of spectrum the system has in the high 
orders of perturbation theory, f.e. whether it splits further or widens. In such an analysis, 
terms occur in $\Sigma$ proportional to $1/T$. The same happens when the system is 
additionally degenerated with respect to spin (i.e. for $g=0, 2$). Having imposed 
the restriction on interactions between particles with the same projection, we thus 
exclude the FQHE. However, our model is reasonable and justified as it leads to no
contradiction and predicts interesting results. 


\section{Acknowledgments}

The persistent and stimulating interest of Yu.~A.~Bychkov is gratefully
appreciated. Helpful discussions with D.~V.~Efremov, A.~Shytov, 
A.~M.~Dyugaev, S.~V.~Iordanskii, and W.~Apel are acknowledged.
The author is grateful to I.~L.~Aleiner for a critical reading of the manuscript
and valuable comments.
This material is based upon work supported by the U.S. Civilian Research
and Development Foundation under Award No.  RP1-273 and by the 
Sonderforschungsbereich 237 ``Unordnung und grosse Fluktuationen''. 

\begin{figure}  

Figure 1 a) Two interacting particles, 
             changing their indices $N_1, N_2$ and momenta 
             $p_1$, $p_2$ to $N_3, N_4$ and $p_3$, $p_4$, respectively.
             The point corresponds to the vertex (\ref{a7}). b) The same
             process for a particular choice of the indices, $N_2=N_0$. 
             The dimple in LL $N^\uparrow_0$, corresponding to the chemical
             potential, depicts the incomplete filling of this level.

Figure 2 a) 
           Second-order diagram for an electron b) 
              Corresponding process for a particular choice of the indices.
                       (c -- d) Same as (a -- b) for a hole.
              Processes with $N_1-N_3=N_4-N_2$ lead to
              singular terms in the self-energy part. 

Figure 3 Second-order processes for the LL $N^\downarrow_0+2$: a) 
              a  ``large'' one, no particle in the dimple in
             the LL $N^\uparrow_0$.  (b -- c) ``Small'' ones, with a particle
              involved in the LL $N^\uparrow_0$.  

Figure 4  Third-order diagram representing interactions of quasiparticles with
          opposite spins. Dotted lines mark interactions, as in  
          Ref. \cite{LUT}. The process evolves with 
          time $t$. Each cut, corresponding to some time between 
          interactions, contributes to the self-energy $\Sigma$ according to Eq.
          (\ref{c1}). Both the incoming and outgoing propogators
          contain external frequency $\omega$ so that the denominator in  Eq. (\ref{c1}) 
          due to cut A-B, $\epsilon^\downarrow_1+\epsilon^\uparrow_2-
          \epsilon^\downarrow_3- \epsilon^\uparrow_4$, contains only energies 
          of LL's $\epsilon_i$. This, however, does not lead to a term in  
          $\Sigma$, proportional to reversed power of temperature $T$ even
          if $\epsilon^\downarrow_1+\epsilon^\uparrow_2=
          \epsilon^\downarrow_3- \epsilon^\uparrow_4$, provided that the
          contact interactions are considered.

Figure 5 A diagram describing consecutive decay of either a particle or a hole.
         The circles around the vertex denote independent laws of 
         momentum conservation. 

Figure 6 a) A ladder graph representing the process of interaction of two
             intermediate states for the second-order diagram (Fig. 2) 
         b) Essential interactions for the LL $N^\downarrow_0$. c) The 
             same for the LL $N^\downarrow_0+1$. In (b -- c)  the interactions 
             are marked as in Ref. \cite{LUT}.

Figure 7 Essential processes: a) for an electron in  LL $N^\uparrow_0+1$;
            b) for a hole in  LL $N^\uparrow_0-1$. 
            
Figure 8 The Dyson equations in the limit $N_0(1-\nu_0)\ll 1$: a) for LL 
                $N^\downarrow_0$; b) for LL $N^\downarrow_0+1$. 
                It is implied that each propagator depends both on    
               $\omega_i$ and $p_i$. The projection of spins are explicitly
               shown in the equation for the self-energy  but omited
               in that of the vertex.
                       
Figure 9 Same as Fig. 8 in the limit $N_0(1-\nu_0) \gg 1$. 

Figure 10 Spectrum in the first regime of renormalization $N^{1/2}_0(1-\nu_0) \ll 1$. 
Left: bare spectrum. Right: renormalized spectrum. Solid lines mark the position of 
unchanged levels. Dashed and dotted lines correspond to the initial position 
of the LL's and new levels arising due to the interaction, respectively.

Figure 11 Spectrum in the second regime of renormalization $1/ \sqrt{N_0} \ll 1-
\nu_0 \ll 1$. Only the relative shift of levels is shown. The meaning of the lines is 
the same as in Fig. 10.

\newpage
\begin{center}
  \unitlength1cm
  \begin{minipage}[t]{12cm}
\hspace*{-2cm}
  \begin{picture}(6,6)
     \put(-3.,-19.5)
    {\includegraphics{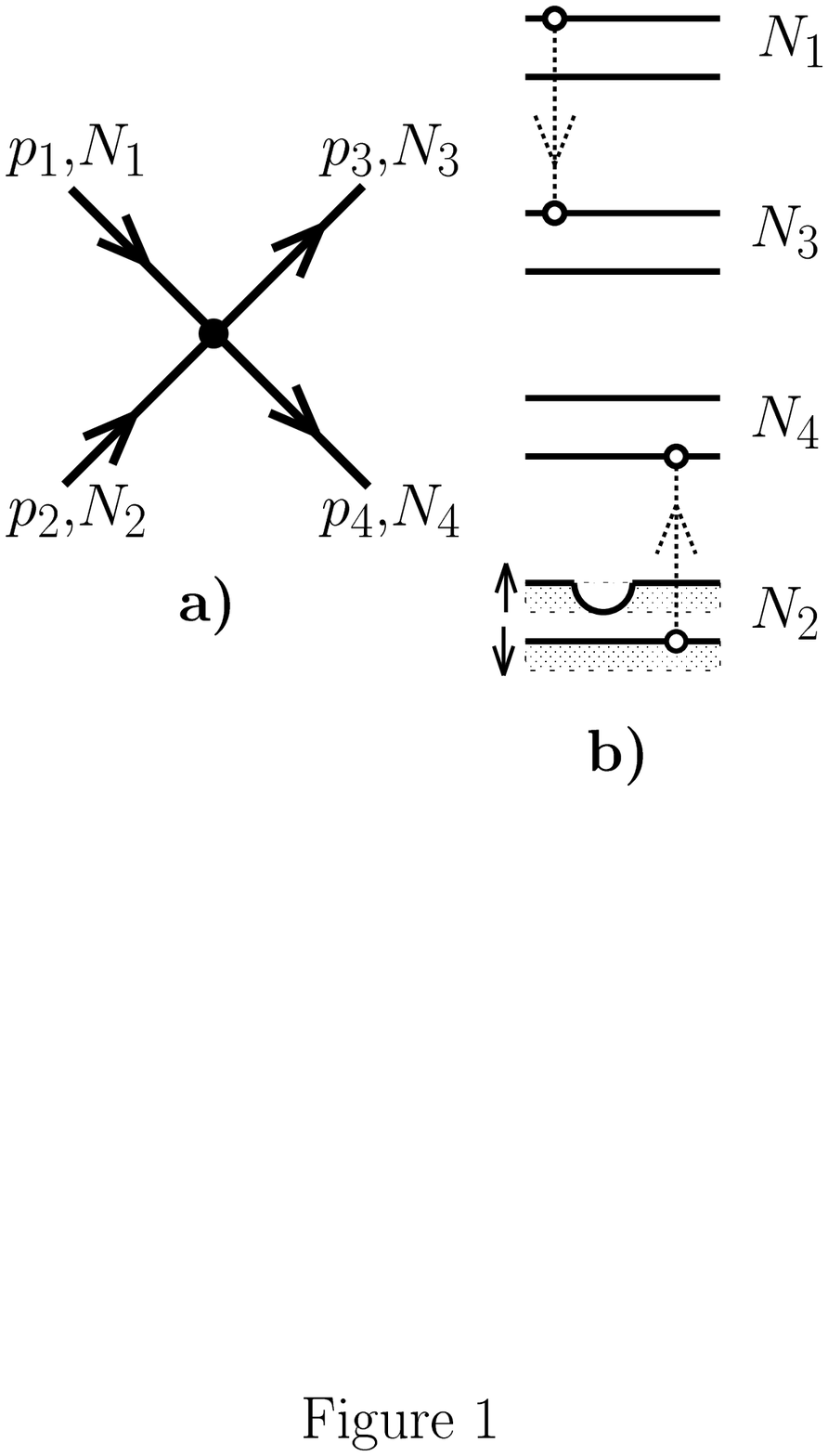}}
\end{picture}  \end{minipage} \end{center} 

\newpage

\begin{center}
  \unitlength1cm
  \begin{minipage}[t]{12cm}
  \begin{picture}(6,6)  
     \put(-4.,-19.) 
    {\includegraphics{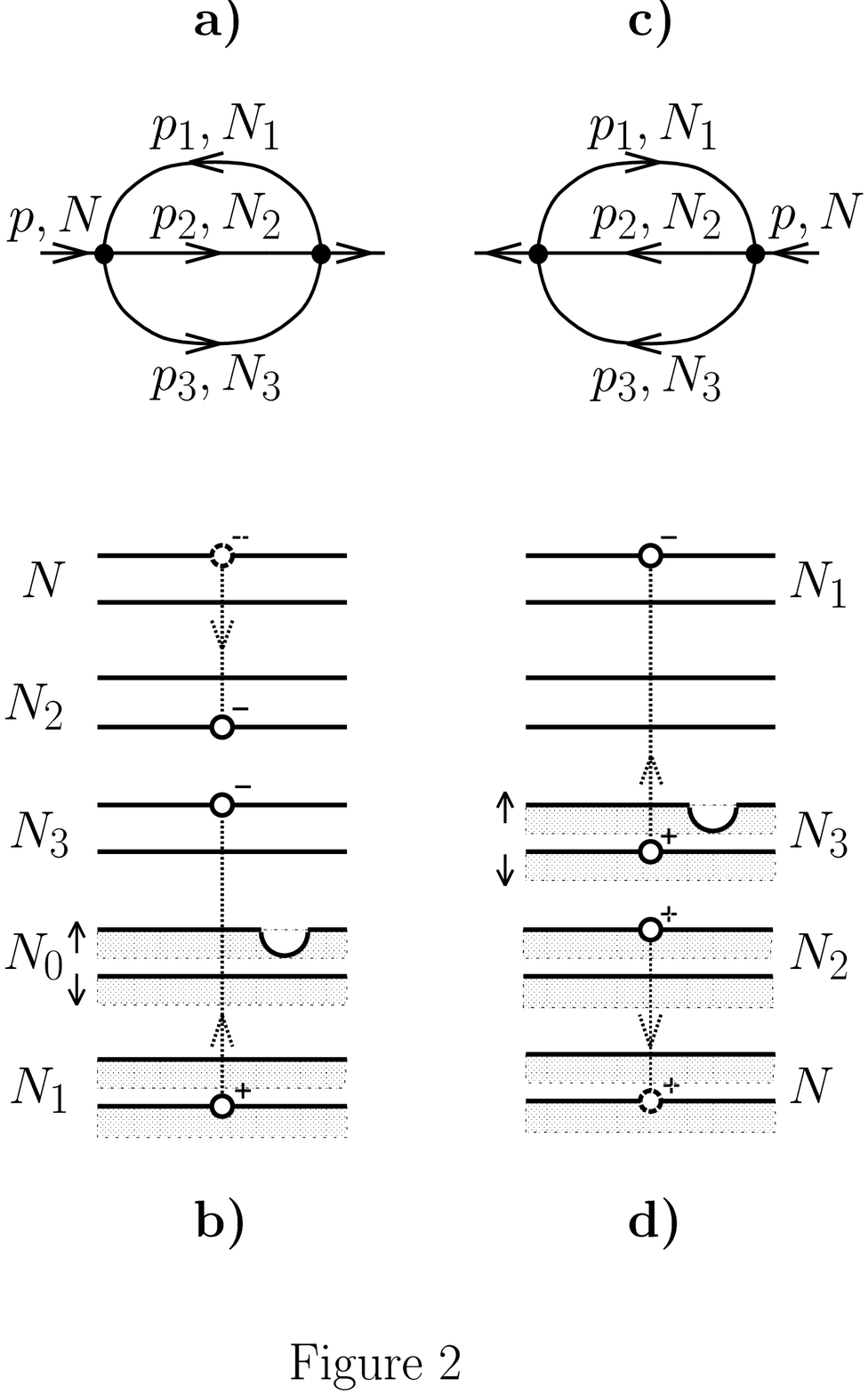}}
 \end{picture}
  \end{minipage}
\end{center}

\newpage

\begin{center}
  \unitlength1cm
  \begin{minipage}[t]{12cm}
  \begin{picture}(0,5)
     \put(-4.,-19.5)
    {\includegraphics{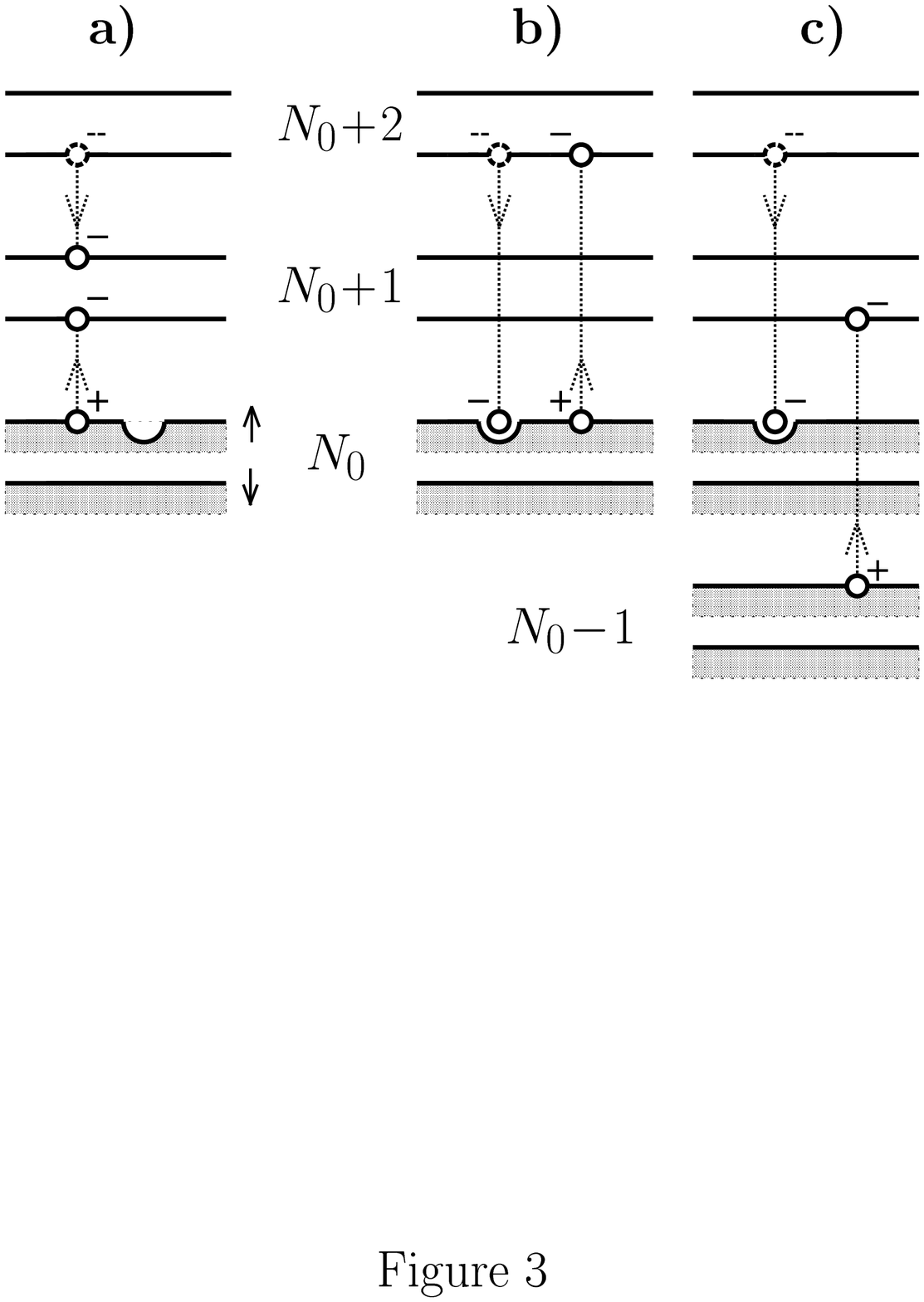}}
 \end{picture}
  \end{minipage}
\end{center}

\newpage

\begin{center}
  \unitlength1cm
  \begin{minipage}[t]{12cm}
  \begin{picture}(0,5)
     \put(-4.,-19.)
    {\includegraphics{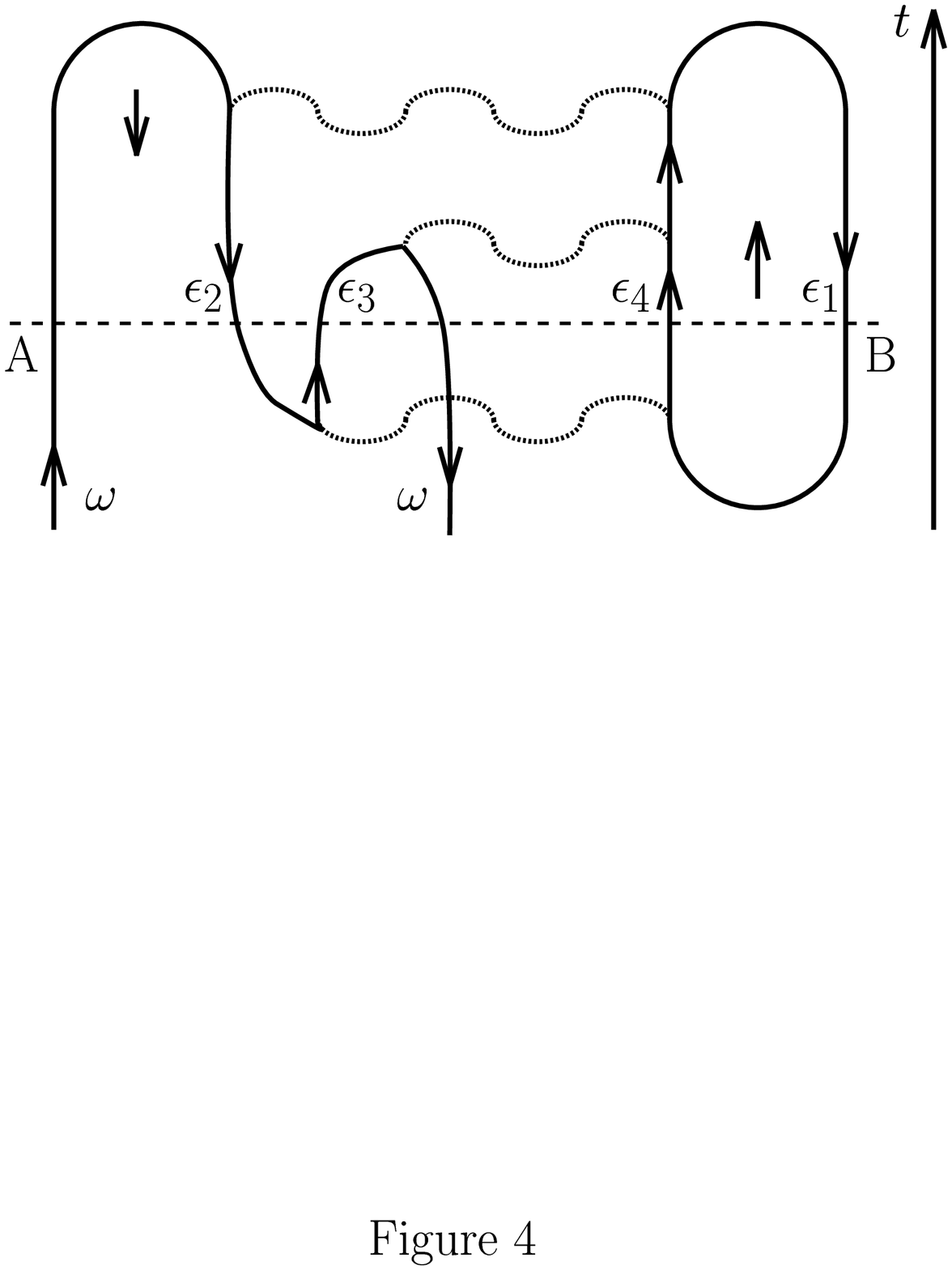}}
 \end{picture}
  \end{minipage}
\end{center}

\newpage

\begin{center}
  \unitlength1cm
  \begin{minipage}[t]{12cm}
  \begin{picture}(0,5)
     \put(-3.5,-17.)
    {\includegraphics{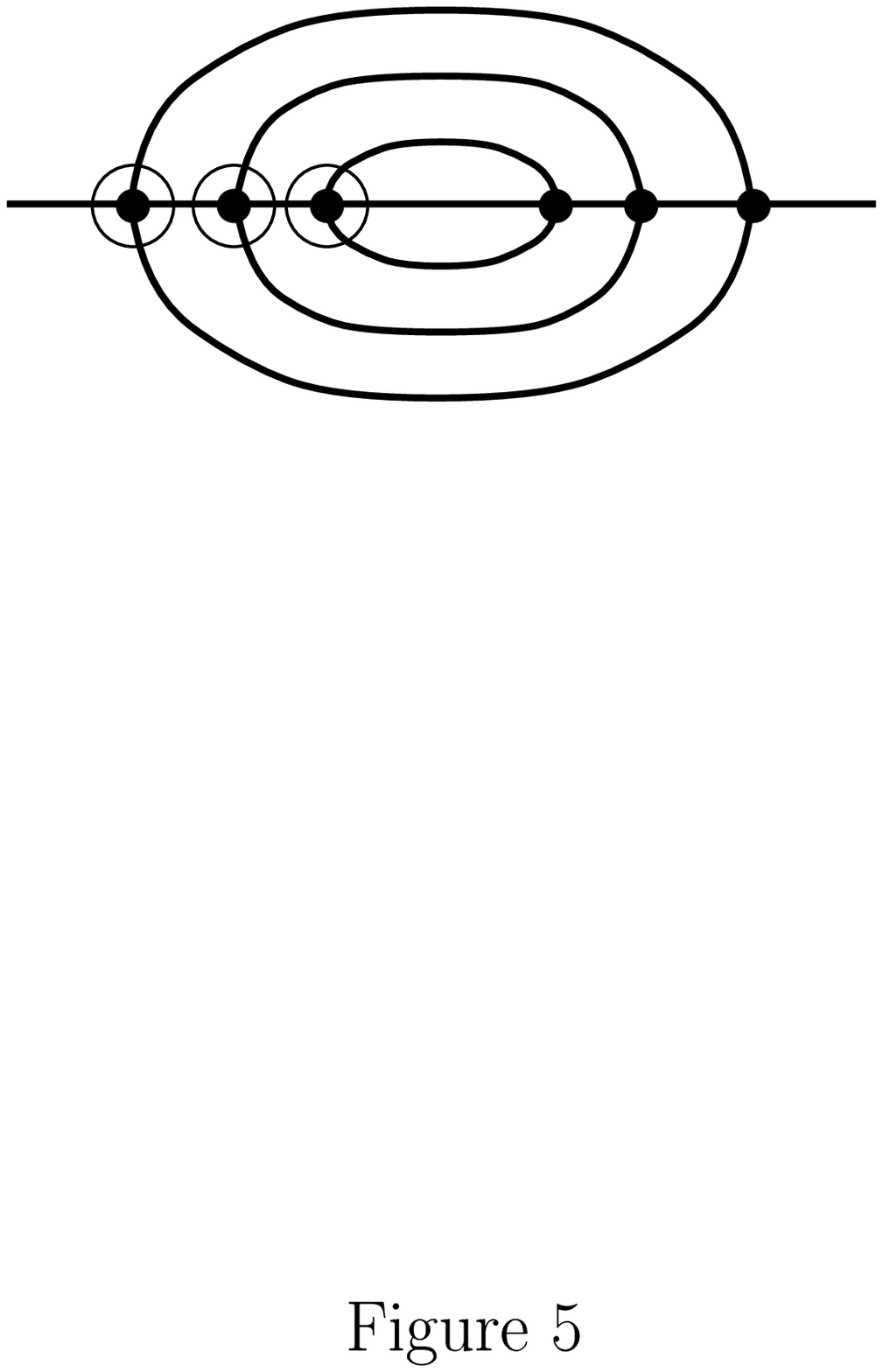}}
 \end{picture}
  \end{minipage}
\end{center}

\newpage

\begin{center}
  \unitlength1cm
  \begin{minipage}[t]{12cm}
  \begin{picture}(0,5)
     \put(-4.,-18.3)
    {\includegraphics{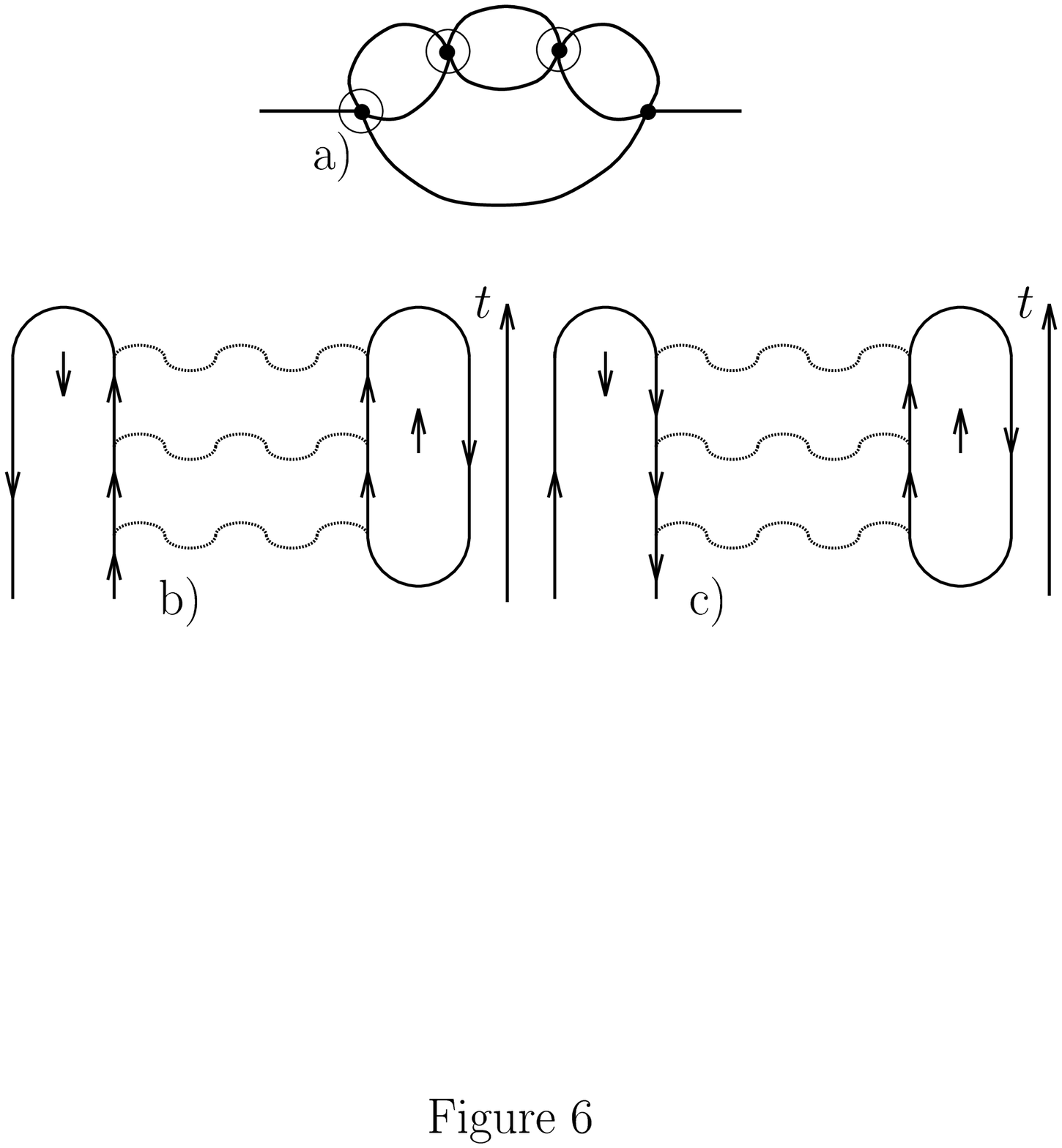}}
 \end{picture}
  \end{minipage}
\end{center}

\newpage

\begin{center}
  \unitlength1cm
  \begin{minipage}[t]{12cm}
  \begin{picture}(0,5)
     \put(-3.5,-18.)
    {\includegraphics{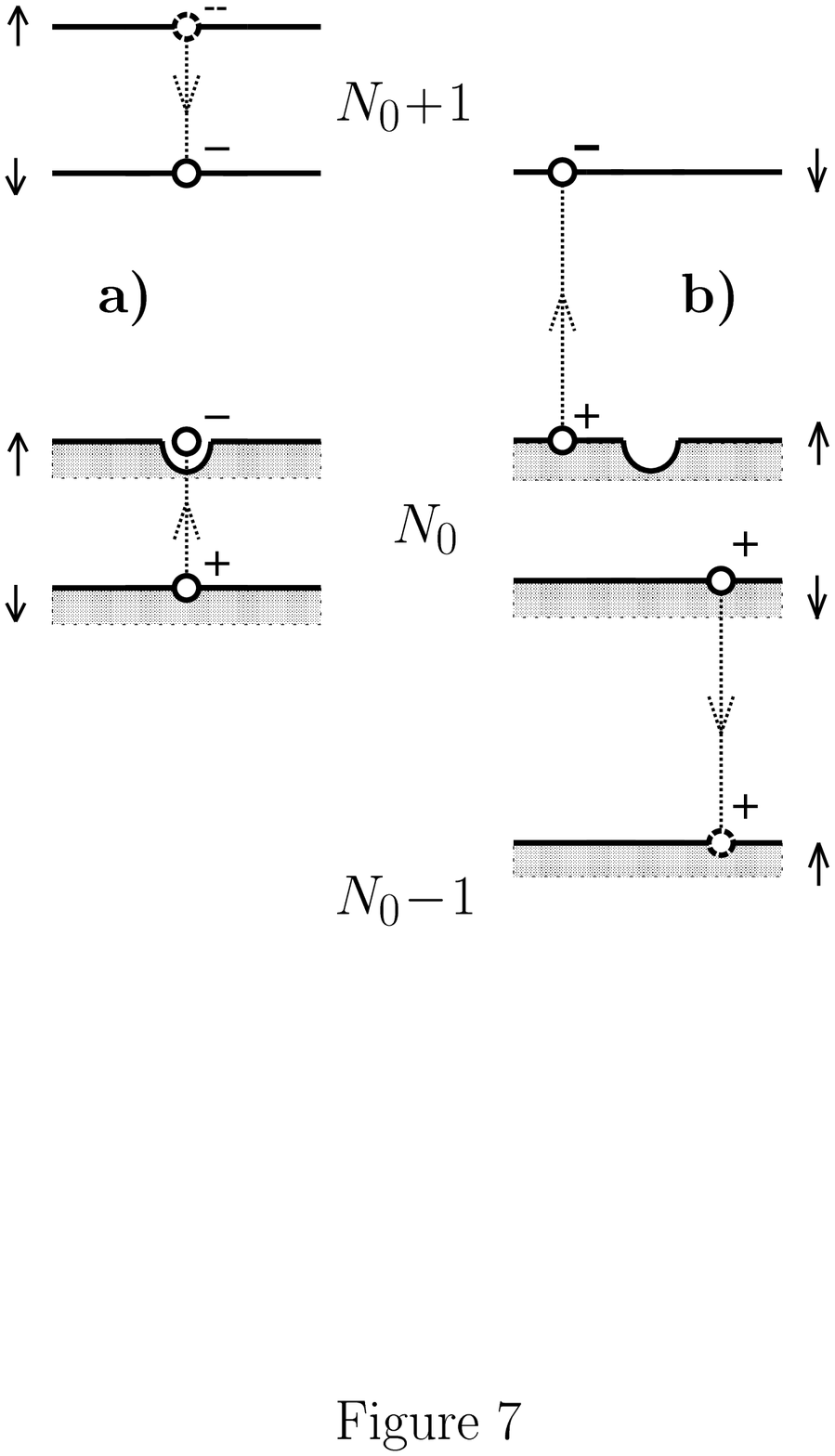}}
 \end{picture}
  \end{minipage}
\end{center}

\newpage

\begin{center}
  \unitlength1cm
  \begin{minipage}[t]{12cm}
  \begin{picture}(0,8)
     \put(-5.3,-20.5)
    {\includegraphics{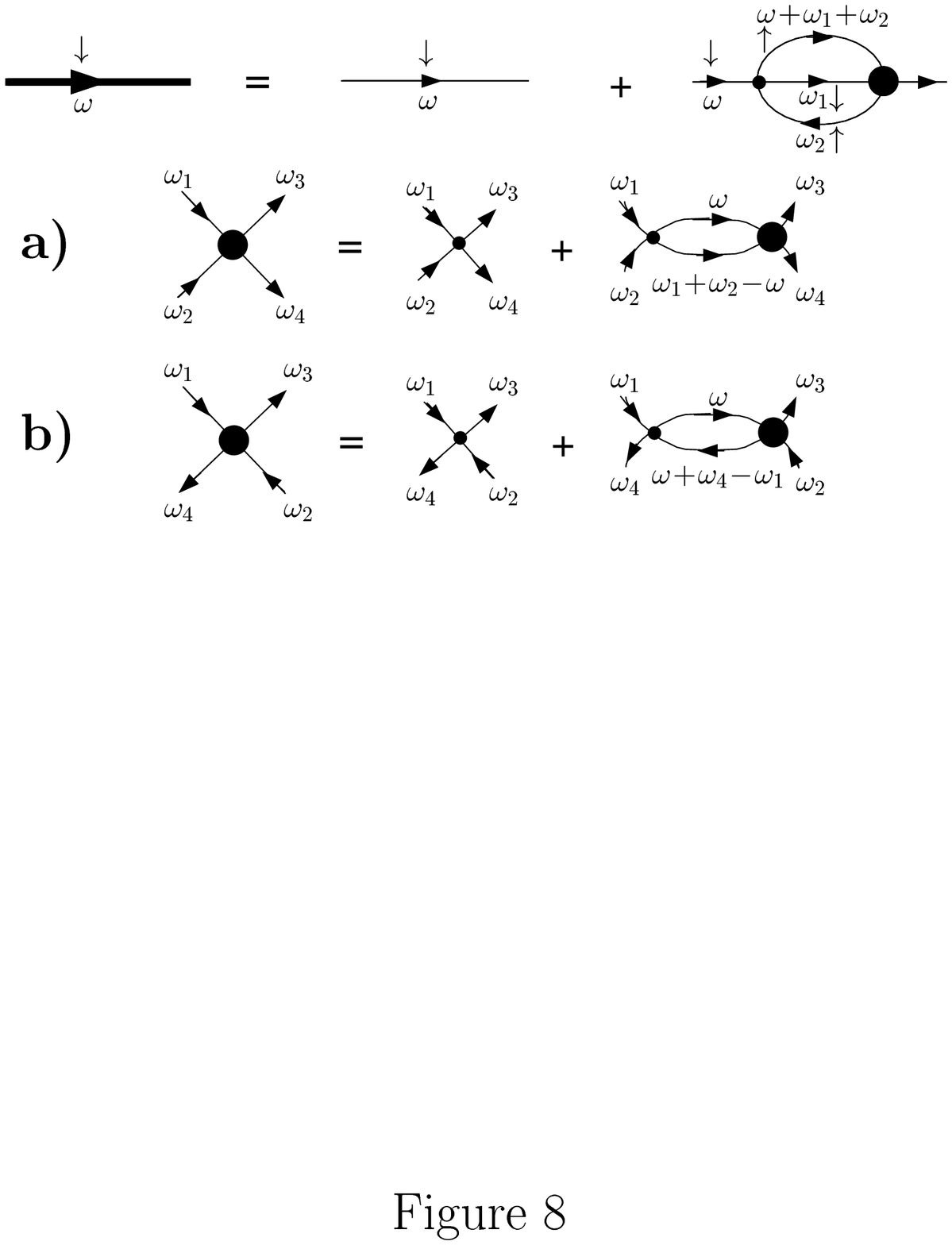}} 
  \end{picture}
  \end{minipage} 
  \end{center}
  \newpage

\begin{center}
  \unitlength1cm
  \begin{minipage}[t]{12cm}
  \begin{picture}(0,8)
     \put(-6.,-20.5)
    {\includegraphics{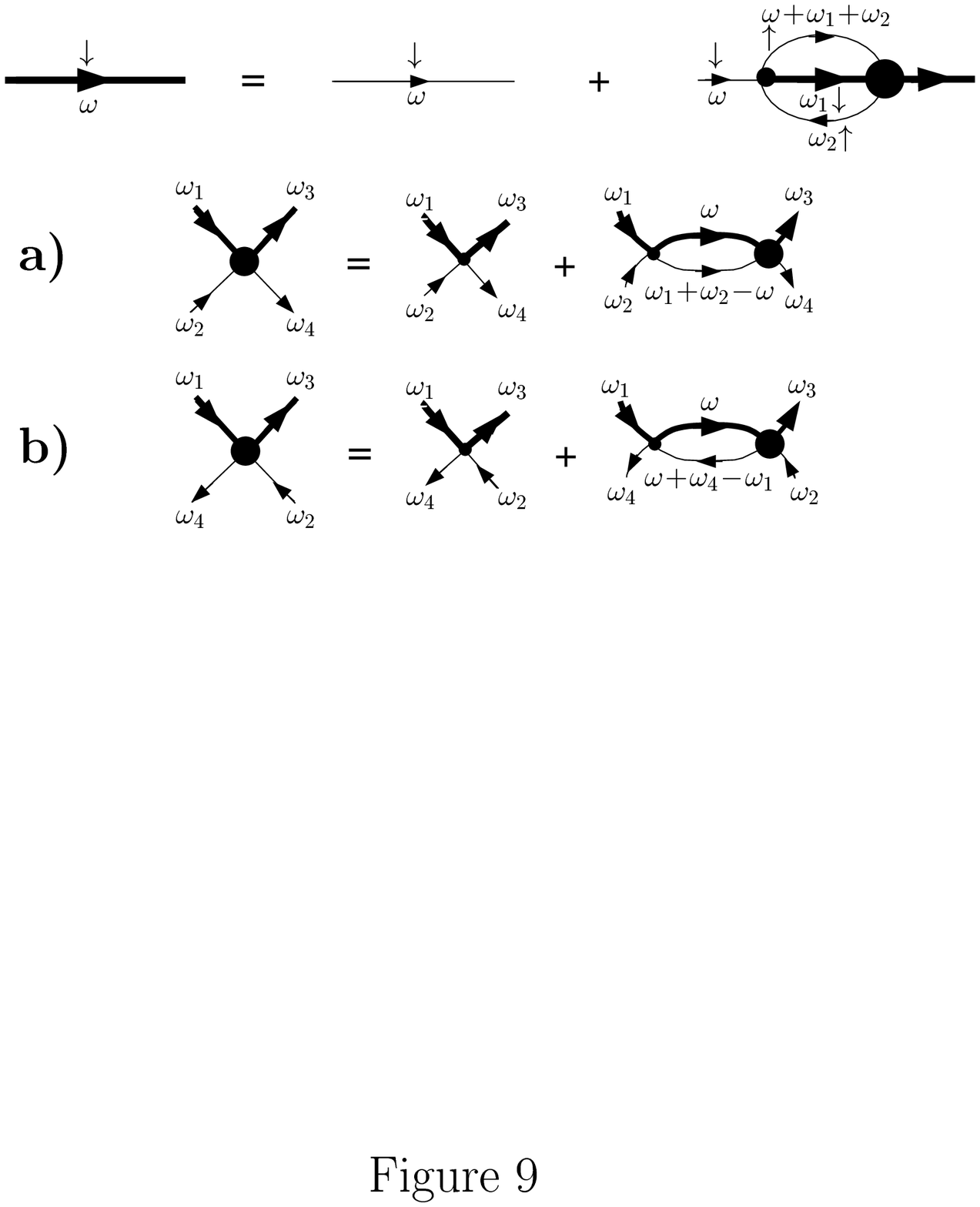}}
 \end{picture}
  \end{minipage}
\end{center}

\newpage

\begin{center}
  \unitlength1cm
  \begin{minipage}[t]{12cm}
\vspace*{-2cm}
  \begin{picture}(0,5)
     \put(-4,-20.5)
    {\includegraphics{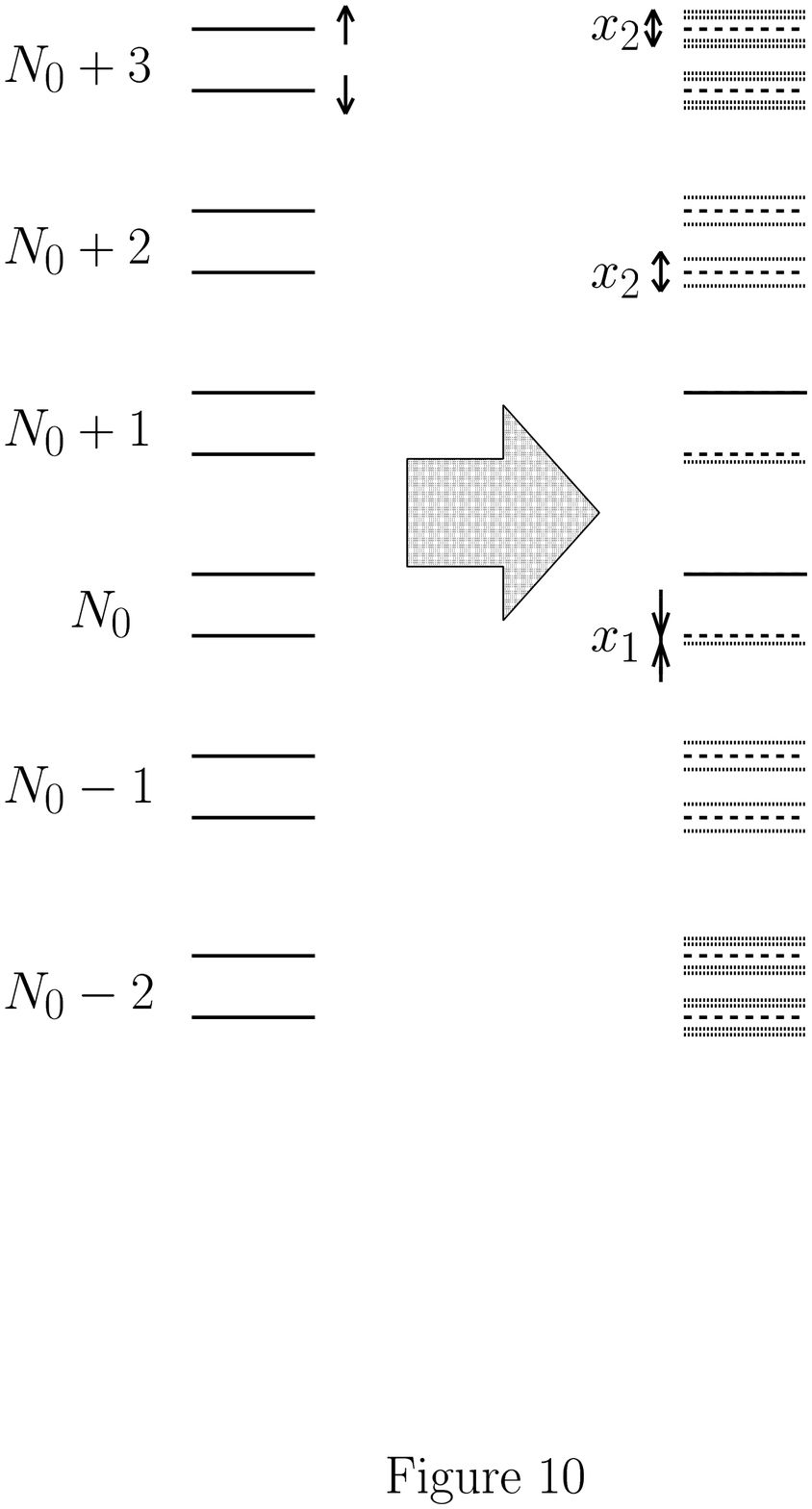}}
 \end{picture}
  \end{minipage}
\end{center}

\newpage

\begin{center}
  \unitlength1cm
  \begin{minipage}[t]{12cm}
\vspace*{-2cm}
  \begin{picture}(0,5)
     \put(-4,-20.5)
    {\includegraphics{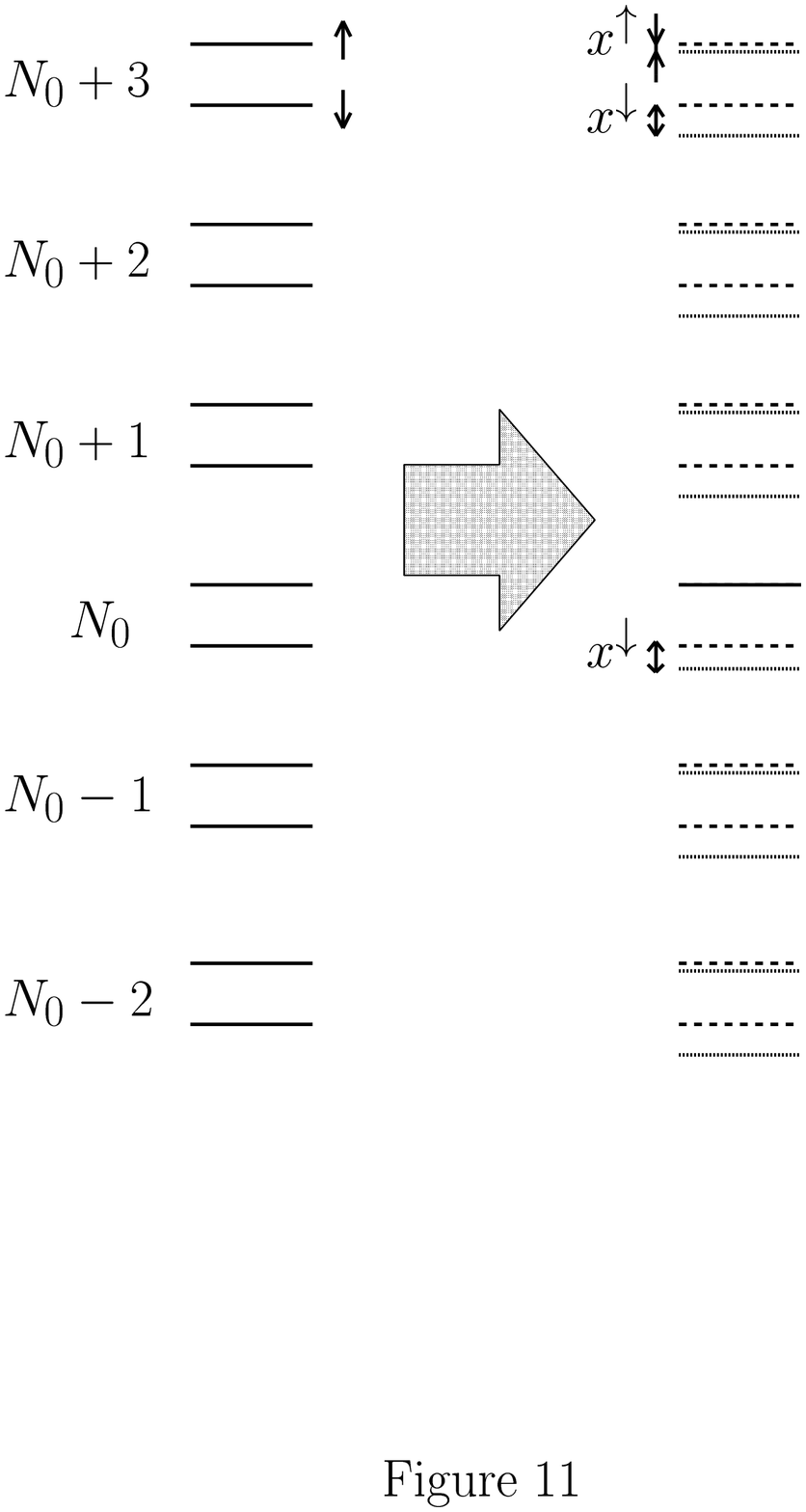}}
  \end{picture}
  \end{minipage}
\end{center}
\end{figure}                                                         
\end{document}